\documentclass[a4paper, 11pt]{article}
\usepackage{jheppub}\makeatletter\gdef\@fpheader{}\makeatother
\usepackage{mathtools, amsfonts, microtype, dcolumn, caption}
\usepackage[section]{placeins}  
\usepackage[bb=dsfontserif]{mathalpha} 
\usepackage[noscale]{youngtab}\Yboxdim 6.5pt\Ylinethick 0.4pt  
\usepackage[dvipsnames]{xcolor}\definecolor{darkgreen}{rgb}{0,0.4,0.2} 
\usepackage{orcidlink}\makeatletter\gdef\@OrigHeightRecip{0.005}\makeatother
\usepackage{tikz} 
\usetikzlibrary{arrows.meta}  
\hypersetup{urlcolor=teal, citecolor=teal, linkcolor=blue}

\graphicspath{{Figures/}} 

\newcommand\mc[1]{\multicolumn{1}{c|}{#1}}
\newcommand{\T}{\mathbb{T}} 
\newcommand{\Z}{\mathbb{Z}}  
\newcommand{\R}{\mathcal{R}}
\newcommand{\U}{\textrm{U}}
\newcommand{\SU}{\textrm{SU}}
\newcommand{\USp}{\textrm{USp}} 
  
\def\K3{\ensuremath{\mathbb{K}3}}

\title{Three-family supersymmetric Pati-Salam models from intersecting D6-branes on rigid cycles}

\author[a]{Adeel Mansha,\,\orcidlink{0000-0002-1183-0355}}
\author[b]{Mudassar Sabir,\,\orcidlink{0000-0002-8551-2608}}
\author[c,d,e]{Tianjun Li,\,\orcidlink{0000-0003-1583-5935}} 
\author[a]{Luyang Wang\,\orcidlink{0000-0001-9400-7331}}  

\affiliation[a]{College of Physics and Optoelectronic Engineering, Shenzhen University, 3688 Nanhai Avenue, Shenzhen, P. R. China}
\affiliation[b]{School of Physics, University of Electronic Science and Technology of China, 2006 Xiyuan Avenue, Chengdu, P. R. China} 
\affiliation[c]{School of Physics, Henan Normal University, 
Xinxiang 453007, P. R. China}
\affiliation[d]{Institute of Theoretical Physics, Chinese Academy of Sciences, 
Beijing 100190, P. R. China}
\affiliation[e]{School of Physical Sciences, University of Chinese Academy of Sciences, No.~19A Yuquan Road, Beijing 100049, P. R. China}

\emailAdd{adeelmansha@alumni.itp.ac.cn}
\emailAdd{mudassar.sabir@uestc.edu.cn}
\emailAdd{tli@itp.ac.cn}  
\emailAdd{wangly@szu.edu.cn}

\keywords{}
\arxivnumber{2505.03664}

\abstract{Intersecting D6-brane models without discrete torsion typically suffer from unstabilized open string moduli, arising from D-brane positions and Wilson lines. These moduli generate additional massless adjoint fields, obstructing the realization of negative beta functions necessary for asymptotic freedom unless they are decoupled around string scale. A viable solution involves utilizing rigid cycles, which eliminate these unwanted adjoint fields. In this work, we for the first time present a class of consistent three-family supersymmetric Pati-Salam models from rigid intersecting D6-branes on the factorizable $\mathbb{T}^6/(\mathbb{Z}_2 \times \mathbb{Z}_2^\prime)$ orientifold with discrete torsion. These models satisfy all the known consistency conditions, including $\mathcal{N}=1$ supersymmetry, K-theory constraints, tadpole cancellation, and recent swampland bounds on the maximal gauge group rank. We provide detailed particle spectra, analyze their phenomenological implications, and discuss the decoupling of exotic states through strong dynamics in the hidden sector.}

\begin{document}
\maketitle  

\section{Introduction}\label{sec:Intro}

A central challenge in string phenomenology is the construction of realistic string vacua that yield low-energy supersymmetric Standard Models (SMs) free from exotic particles, while simultaneously stabilizing all the moduli fields. Such models can be connected to the observable particle physics through renormalization group equations evolution, allowing for experimental probes at the Large Hadron Collider and future collider experiments. Significant progress has been made in this direction within Type IIA string theory compactified on the $\T^6/(\Z_2 \times \Z_2)$ orientifold, whereby the semi-realistic supersymmetric Standard-like Models have been extensively developed~\cite{Cvetic:2001tj, Cvetic:2001nr, Cvetic:2002pj, Cvetic:2004ui, Blumenhagen:2005mu, Chen:2005mj, Chen:2006gd, Chen:2006ip, Chen:2007px, Chen:2007zu, He:2021kbj, He:2021gug, Li:2022cqk, Sabir:2022hko, Li:2019nvi, Li:2021pxo, Mansha:2022pnd, Mansha:2023kwq, Sabir:2024cgt, Sabir:2024jsx, Sabir:2024mfv, Mansha:2024yqz}. Among these constructions, Pati-Salam-like models are particularly compelling, as they can naturally accommodate all the Yukawa couplings~\cite{Cvetic:2004ui}. In fact, the first systematic realization of three-family supersymmetric Pati-Salam models was presented in~\cite{Cvetic:2004ui}. Such kind of models can also be constructed in the heterotic string framework~\cite{Assel:2010wj, Faraggi:2020wld}. The SM fermion masses and mixings have been explicitly explained in certain intersecting D-brane models~\cite{Sabir:2022hko, Sabir:2024cgt}. However, a common feature of these semi-realistic constructions is the presence of three adjoint chiral multiplets for each $\U(N)$ gauge symmetry, arising due to the non-rigidity of the D6-branes. On the other hand, while models based on rigid intersecting D6-branes eliminate such adjoint moduli, they remain far from realistic in terms of reproducing the full chiral and Higgs spectrum~\cite{Blumenhagen:2005tn, Forste:2008ex, Blumenhagen:2002wn, Ibanez:2008my, Blumenhagen:2006xt, Ibanez:2006da, Abel:2006yk, Blumenhagen:2007zk}. Therefore, constructing three-family $\mathcal{N}=1$ supersymmetric Pati-Salam models using rigid intersecting D6-branes continues to be a major challenge in string model building. 

In ref.~\cite{Blumenhagen:2005tn}, it was first demonstrated that global $\mathcal{N}=1$ chiral models can be constructed using rigid intersecting D6-branes on a factorizable $\T^6/(\Z_2 \times \Z_2^\prime)$ orientifold with discrete torsion. However, engineering consistent three-family models in this setup with factorizable tori has proven to be particularly challenging. In the case of non-rigid branes, we need at least one torus to be tilted to generate three-generation models on the mirror orientifold $\T^6/(\Z_2\times \Z_2)$ \cite{Cvetic:2001tj, Cvetic:2001nr, Cvetic:2002pj, Cvetic:2004ui, Blumenhagen:2005mu, Chen:2005mj, Chen:2006gd, Chen:2006ip, Chen:2007px, Chen:2007zu, He:2021kbj, He:2021gug, Li:2022cqk, Sabir:2022hko, Li:2019nvi, Li:2021pxo, Mansha:2022pnd, Mansha:2023kwq, Sabir:2024cgt, Sabir:2024jsx, Mansha:2024yqz}. In the case of rigid 3-cycles, however, no consistent three-family models incorporating tilted tori have been reported. A notable example of a three-generation model based on rigid D6-branes was presented in ref.~\cite{Forste:2008ex}, but it was realized on non-factorizable tori. The corresponding model with factorizable tori fails to satisfy the $\mathcal{N}=1$ supersymmetry conditions. Moreover, that construction featured a significantly enlarged gauge group, exceeding phenomenological expectations.

Recent insights from the swampland program~\cite{Brennan:2017rbf} impose additional constraints on string model building by bounding the maximal rank of the gauge group in consistent quantum gravity theories. For theories with 32 supercharges, supersymmetry is strong enough to completely fix all possible configurations. For the next class, with 16 supercharges, it has been shown in \cite{Kim:2019ths} that in $d = 10$ and in the non-chiral cases, the possibilities are finite because the rank of the gauge group must satisfy $r(G) \leq 26 - d$. For six-dimensional theories with eight supercharges, it has been established recently in \cite{Kim:2024hxe} that the number of tensor multiplets is bounded by $T \leq 193$, and the rank of the gauge group is constrained by $r(V) \leq 480 $. However, for lesser supersymmetry, say for $\mathcal{N}=1$ Pati-Salam models the maximal rank was found to be 138 in \cite{Loges:2022mao}. For reviews on swampland program, please see \cite{Brennan:2017rbf, Agmon:2022thq, Palti:2019pca, vanBeest:2021lhn, VanRiet:2023pnx}.   

In this paper, we for the first time construct a class of consistent three-family $\mathcal{N}=1$ supersymmetric Pati-Salam models with factorizable tori, based on rigid intersecting D6-branes on a factorizable $\mathbb{T}^6/(\mathbb{Z}_2 \times \mathbb{Z}_2')$ orientifold with discrete torsion \footnote{A representative example from this class is also discussed in our companion letter~\cite{Mansha:2025gvr}.}. This background naturally accommodates the rigid three-cycles that can be wrapped by D6-branes, enabling the construction of chiral models in which moduli associated with D-brane positions are absent, or more precisely, such would-be moduli acquire masses at the string scale. The construction is particularly tractable, as all the three two-tori are taken to be rectangular, each containing four fixed points. Consequently, rigid D6-branes are restricted to pass through specific pairs of fixed points on each two-torus. A notable feature of this setup is the absence of adjoint matter, which is highly advantageous for realistic model building, as it eliminates unwanted scalar fields.

The construction utilizes semi-rigid and non-rigid hidden sector branes to ensure the complete cancellation of twisted and untwisted RR tadpoles, while simultaneously maintaining an odd number of chiral families. The presence of the massless Higgs pairs for the Pati-Salam gauge symmetry breaking is achieved through the controlled recombination of hidden sector branes with the stack supporting the $\SU(2)_R$ gauge symmetry, leading to realistic symmetry-breaking patterns without introducing exotic chiral matters.

The paper is organized as follows. In Section~\ref{sec:model-building}, we review the model-building rules used to construct three-family Pati-Salam models from configurations of intersecting rigid D6-branes. We also revise the calculation of intersection numbers in this setting, along with the associated consistency conditions arising from $\mathcal{N}=1$ supersymmetry, K-theory constraints, and the cancellation of RR tadpoles. In Section~\ref{sec:models}, after outlining the model building strategy, we present a number of consistent three-family models with detailed particle spectra, and discuss the phenomenological aspects of these models, particularly focusing on the decoupling of exotic particles from the low-energy spectrum via strong dynamics. In Section~\ref{sec:asymptoticfreedom}, we discuss the realization of asymptotic freedom in all models by evaluating the beta functions, which provides yet another consistency check on the viability of the constructed models. Finally, we conclude in Section~\ref{sec:conclusion}.

\section{Pati-Salam model building on rigid cycles}\label{sec:model-building}
We consider the type IIA $\T^6/(\Z_2\times \Z_2^\prime)$ orientifold with the \textit{discrete torsion} turned on. Here, $\T^6$ is a product of three 2-tori with the orbifold group $(\mathbb{Z}_2\times \mathbb{Z}_2')$ having the generators $\theta$ and $\omega$ respectively associated with the twist vectors $(1/2,-1/2,0)$ and $(0,1/2,-1/2)$ such that their action on complex coordinates $z_i$ is given by,
\begin{eqnarray}
	& \theta: & (z_1,z_2,z_3) \to (-z_1,-z_2,z_3), \nonumber \\
	& \omega: & (z_1,z_2,z_3) \to (z_1,-z_2,-z_3). \label{orbifold}
\end{eqnarray}
Orientifold projection is the gauged $\Omega \R$ symmetry, where $\Omega$ is world-sheet parity that interchanges the left- and right-moving sectors of a closed string and swaps the two ends of an open string while $\R$ acts as complex conjugation on coordinates $z^I$ as,
\begin{equation}
\begin{array}{ccccc}
\Omega :      & (\sigma_1, \sigma_2) & \mapsto & (2\pi - \sigma_1, \sigma_2) & \text{(Closed)} \\
              & (\tau, \sigma)       & \mapsto & (\tau, \pi - \sigma)        & \text{(Open)}   \\
\mathcal{R} : &       z_i            & \mapsto & \overline{z}_i.             &
\end{array}
\end{equation}
This results in four different kinds of orientifold 6-planes (O6-planes) corresponding to $\Omega \R$, $\Omega \R\theta$, $\Omega \R\omega$, and $\Omega \R\theta\omega$ respectively. These orientifold projections are only consistent with either the rectangular or the tilted complex structures of the factorized 2-tori. Denoting the wrapping numbers for the rectangular and tilted tori as $n_a^I[a^I]+m_a^I[b^I]$ and $n_a^I[a'^I]+m_a^I[b^I]$ respectively, where $[a'^I]=[a^I]+\frac{1}{2}[b^I]$. Then a generic 1-cycle $(n_a^I,l_a^I)$ satisfies $l_{a}^I\equiv m_{a}^I$ for the rectangular 2-torus and $l_{a}^I\equiv 2\tilde{m}_{a}^I=2m_{a}^I+n_{a}^I$ for the tilted 2-torus such that $l_a^I-n_a^I$ is even for the tilted tori. The two different basis $(n^i,m^i)$ and $(n^i,l^i)$ are related as,
\begin{align}\label{basis-l-m}
	l^i & = 2^{\beta_i} (m^i + \frac{\beta_i}{2} n^i), \qquad \beta_i = \begin{cases} 
	0 ~ & \mathrm{rectangular}~\T^2,                                                  \\ 
	1 ~ & \mathrm{tilted}~\T^2. \end{cases}                                           
\end{align}  

In contrast to the $\mathbb{Z}_2 \times \mathbb{Z}_2$ orbifold without discrete torsion, where all O6-planes are of type O6$^{(-,-)}$, the inclusion of discrete torsion ($\eta = -1$) imposes consistency conditions on the crosscap states, requiring the presence of an odd number of exotic O6$^{(+,+)}$-planes. This condition arises from the compatibility of the relations
\begin{align}
	\langle \Omega \mathcal{R} | e^{-l \mathcal{H}_{\text{cl}}} | \Omega \mathcal{R} \omega \rangle               & = \text{Tr}_{\omega} \left( \Omega \mathcal{R}\, e^{-2\pi t H} \right), \nonumber \\
	\langle \Omega \mathcal{R} \theta | e^{-l \mathcal{H}_{\text{cl}}} | \Omega \mathcal{R} \theta \omega \rangle & = \text{Tr}_{\omega} \left( \Omega \mathcal{R} \theta\, e^{-2\pi t H} \right),  
\end{align}
which, along with the choice of discrete torsion $\eta = \pm 1$ in the closed string sector, enforces the constraint
\begin{align}
	\eta_{\Omega \mathcal{R}}\, \eta_{\Omega \mathcal{R}\theta}\, \eta_{\Omega \mathcal{R}\omega}\, \eta_{\Omega \mathcal{R}\theta\omega} = \eta \,. \label{opsigns} 
\end{align} 
Clearly, the choice of discrete torsion, $\eta = -1 $, requires an odd number of O6$^{(+,+)}$-planes~\cite{Blumenhagen:2005tn}.

The discrete torsion gives rise to three $\Z_2$-twisted sectors with sixteen fixed points each. The 3-cycles of $\T^6$ will be inherited by the orbifold quotient. However, in order to deal with 3-cycles on orbifold spaces we have to carefully distinguish between 3-cycles on the covering space and 3-cycles on the actual orbifold. In the particular case at hand, under the action of $\Z_2\times \Z_2^{\prime}$ a factorizable 3-cycle on $\T^6$ has 3 images, all of them with the same wrapping numbers as the initial 3-cycle. Therefore, a 3-cycle in the {\it bulk} of the orbifold space can be identified with $[\Pi_a^B] = 4\, [\Pi_a^{\T^6}]$. Computing the intersection number we get
\begin{align} \label{prod}
	[\Pi_a^B] \circ [\Pi_b^B] & = 4\, [\Pi_a^{\T^6}] \circ [\Pi_b^{\T^6}] = 4 \prod_{I=1}^3 (n^I_a\, \widetilde m^I_b - \widetilde m^I_a\, n^I_b) \nonumber \\
	                          & = 4\cdot2^{-k}\prod_{I=1}^3 (n^I_a\, l^I_b - l^I_a\, n^I_b),  \qquad \because k = \sum_{I=1}^3\beta^I,                       
\end{align}
where we have identified the intersection points related by the $\Z_2\times \Z_2^{\prime}$ action. 

In addition to the untwisted cycles we have 32 independent collapsed three-cycles for each of the three twisted sectors, $\theta$, $\omega$, and $\theta\omega$. Let us first consider the $\theta$-twisted sector. We denote the 16 fixed points on $(\T^2_1 \times \T^2_2)/\Z_2$ by $[e^\theta_{ij}]$, with $i,j\in\{1,2,3,4\}$. After blowing up the orbifold singularities, these become two-cycles with the topology of $\mathbb{S}^2$. Each such four-dimensional $\T^4/\Z_2$ is an orbifold of \K3 before taking the other elements of the orientifold group into account.

With our choice of discrete torsion, these two-cycles are combined with a one-cycle  $(n^3,\widetilde m^3)$ of $\T^2_3$, in order to form a three-cycle in the $\theta$-twisted sector. Let us denote a basis of such twisted three-cycles as,
\begin{align}
	[\alpha^\theta_{ij,\,n}] = 2\, [e^\theta_{ij}]\otimes  [a^{3}]\ , \qquad	[\alpha^\theta_{ij,\,m}] = 2\, [e^\theta_{ij}]\otimes [b^{3}]\ , 
\end{align}
where the factor of two is due to the action of the second $\Z_2$. Analogously, we define the basic twisted three-cycles in the $\omega$ and $\theta\omega$ twisted sectors as,
\begin{align} 
	[\alpha^{\omega}_{ij,\,n}]       & = 2\, [e^{\omega}_{ij}]\otimes [a^1]\ , ~~\qquad  [\alpha^{\omega}_{ij,\,m}] = 2\, [e^{\omega}_{ij}]\otimes [b^1] , \nonumber       \\
	[\alpha^{\theta\omega}_{ij,\,n}] & = 2\, [e^{\theta\omega}_{ij}]\otimes [a^2]\ ,\qquad  [\alpha^{\theta\omega}_{ij,\,m}] = 2\, [e^{\theta\omega}_{ij}]	\otimes [b^2] . 
\end{align}
The intersection number between a pair of such cycles is easy to compute knowing that the collapsed two-cycles of the \K3-orbifold have self-intersection number $[e_{ij}] \circ [e_{kl}] = -2 \delta_{ik} \delta_{jl}$ in each twisted sector, and that two-cycles of different twisted sectors do not intersect. For the three-cycles $[\Pi^g_{ij,\,a}]= n_a^{I_g} [\alpha_{ij,\,n}] +  \widetilde m_a^{I_g} [\alpha_{ij,\, m}]$ and $[\Pi^h_{kl,\,b}]= n_b^{I_h}[\alpha_{kl,\,n}] + \widetilde m_b^{I_h} [\alpha_{kl,\, m}]$, with $g,h = \theta, \omega, \theta\omega$, we find,
\begin{align}\label{inttwist}
[\Pi^g_{ij,\,a}] \circ [\Pi^h_{kl,\,b}]\, & = 4\,\delta_{ik} \delta_{jl} \delta^{gh} \,(n_a^{I_g}\, \widetilde m_b^{I_g} - \widetilde m_a^{I_g}\, n_b^{I_g})\nonumber \\
	                                          & = 4\, \delta_{ik} \delta_{jl} \delta^{gh} 2^{-\beta^g}\,(n_a^{I_g}\, l_b^{I_g} - l_a^{I_g}\, n_b^{I_g}), \qquad  \because \widetilde m_a^{I_g} \equiv 2^{-\beta^g} l_a^{I_g} \,.
\end{align}
In this notation, for the sectors twisted by $g = \theta, \omega,\theta\omega$ one has $I_g = 3,1,2$, respectively.

To build rigid D6-branes, one considers fractional D6-branes charged under all three twisted sectors of the orbifold. Let us start with a factorizable three-cycle, described by three pairs of wrapping numbers  $(n_a^I, \widetilde m_a^I)$. A fractional D6-brane should be invariant under the orbifold action, and hence it must run through four fixed points for each twisted sector. We denote such a set of four fixed points of the element $g$ (each labeled by a pair $(i,j)$) as $S_g^a$. It is easy to characterize the sets of pairs of fixed points $S^a_g$ in terms of the wrapping numbers $(n,m)$ as shown in table~\ref{fixed}.

\begin{table}[t]
	\renewcommand{\arraystretch}{1.4}\centering
	\begin{tabular}{|c|c|c|}
		\hline
		$(n_a, m_a)$ & Fixed points $S^a_g$   \\
		\hline \hline
		(odd, odd)   & $\{1,4\}$ or $\{2,3\}$ \\
		(odd, even)  & $\{1,3\}$ or $\{2,4\}$ \\
		(even, odd)  & $\{1,2\}$ or $\{3,4\}$ \\
		\hline
	\end{tabular}
	\caption{Fixed points of a 1-cycle on a $\T^2/\Z_2$ in terms of its wrapping numbers.}
	\label{fixed}
\end{table}

Then the entire three-cycle that such a fractional D-brane is wrapping is of the form
\begin{align} \label{rigid}
\Pi_a\, &= \frac{1}{4}\, \Pi^B_a  + \frac{1}{4} \sum_{(i,j) \in S_\theta^a} \epsilon^\theta_{a,ij}\,\Pi^\theta_{ij,\,a} + \frac{1}{4}  \sum_{(j,k)\in S_{\omega}^a} \epsilon^{\omega}_{a,jk}\, \Pi^{\omega}_{jk,\,a} + \frac{1}{4} \sum_{(i,k)\in S_{\theta\omega}^a} \epsilon^{\theta\omega}_{a,ik}\, \Pi^{\theta\omega}_{ik,\,a}\ , 
\end{align}
where the signs $\epsilon^\theta_{a,ij},\,\epsilon^{\omega}_{a,jk},\,\epsilon^{\theta\omega}_{a,ik}\,=\,\pm 1$ define the charge of the fractional brane $a$ with respect to the massless fields living at the various fixed points. Geometrically, these numbers indicate the two possible orientations with which the brane can wrap around the blown-up $\mathbb{S}^2$. Clearly, only those fixed points appear in (\ref{rigid}), which the D6-brane is passing through. Since the brane is stuck at the orbifold fixed points in all three $\T_I^2$, there are no adjoint scalars appearing in the massless spectrum. It is \textit{necessary} to use $m$ (and not $l$) to calculate the fixed points in the standard notation of \eqref{rigid}, while $\widetilde{m}$ (or equivalently $2^{-\beta}l$) is useful in the calculation of intersection numbers, tadpole cancelation and the supersymmetry conditions \cite{Forste:2010gw}. 
 
\subsection{Spectrum from rigid branes (\texorpdfstring{$\eta =-1$}{η = -1})}
\begin{table}[t]
	\centering\renewcommand{\arraystretch}{1.4}
	$\begin{array}{|c|c|}
		\hline
		\text{Representation}            & \text{Multiplicity}                                                       \\
		\hline\hline
		(\yng(1)_a,\overline{\yng(1)}_b) & \Pi_a\circ \Pi_{b}                                                        \\
		(\yng(1)_a,\yng(1)_b)            & \Pi_a\circ \Pi'_{b}                                                       \\
		\yng(1,1)_a                      & \frac{1}{2}\left(\Pi_a\circ \Pi'_a + \Pi_{a}\circ \Pi_{\text{O}6}\right)  \\
		\yng(2)_a                        & \frac{1}{2}\left(\Pi_a\circ \Pi'_a - \Pi_{a} \circ \Pi_{\text{O}6}\right) \\
		\hline
	\end{array}$
	\caption{Chiral spectrum for intersecting D6-branes}
	\label{tcs}
\end{table}
The details of the orientifold action prove quite important in order to work out the massless spectrum of the low energy theory. In particular, they are essential in order to compute the chiral matter content which, while still described in terms of intersection numbers between 3-cycles, now includes chiral fermions transforming in the symmetric and anti-symmetric representations of $\U(N)$ gauge groups. For simplicity, let us first consider D6-branes wrapping 3-cycles not invariant under $\R$, so that the gauge group is of the form $\prod_a \U(N_a)$. In this case, we can apply a general rule for determining the massless left-handed chiral spectrum in terms of 3-cycles intersection numbers, as presented in table \ref{tcs}.

Given \eqref{prod} and \eqref{inttwist}, it is now easy to compute the intersection number between two rigid D6-branes of the form (\ref{rigid}).
\begin{align}\label{Iab}
	\Pi_a^F \circ \Pi_b^F & = \frac{1}{4}\Bigg(2^{-k}\prod_I (n^I_a\,l_b^I - l_a^I\,n^I_b) + \sum _{g\in \{3,1,2\}} \delta_{ab}^g\,2^{-\beta^{g}}(n^g_a\,l_b^g - l_a^g\,n^g_b)\Bigg), 
\end{align} 
where $k$ is the number of tilted tori and $\delta_{ab}^g$ is the number of common fixed points where the branes $a$ and $b$ intersect for each twisted sector $g \in \theta, \omega,\theta\omega$ and can be read from \eqref{inttwist} as,
\begin{align}\label{eq:deltaij}
	\delta_{a b}^g\equiv \delta^g(\alpha^a_{ij}, \alpha^b_{kl}) & = \delta^g_{S^a_{i},S^b_{k}} \delta^g_{S^a_{j},S^b_{l}}. 
\end{align}
Assuming that every fractional brane intersects the origin the above relation simplifies as,
\begin{align}
	\delta_{ab}^g & = \sum _{i=1}^2 \sum _{j=1}^2 \sum _{k=1}^2 \sum _{l=1}^2 \delta^g_{S^a_{1,i},S^b_{1,k}} \delta^g_{S^a_{2,j},S^b_{2,l}}~. \label{eq:delta} 
\end{align}

\subsubsection{Orientifold action}  

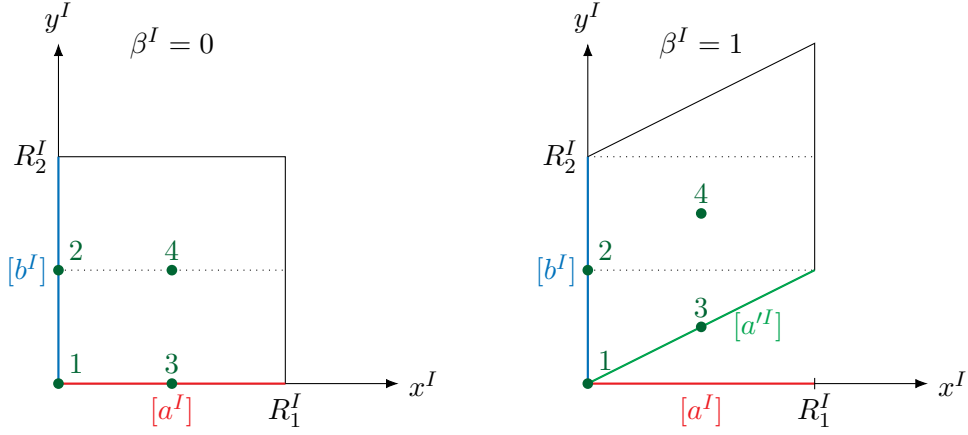
\begin{figure}[t]
	\centering
	\begin{tikzpicture}[x=3cm, y=3cm, scale=1] 
		\begin{scope}
			\draw[-Latex] (0,0) -- (1.5,0) node[right] {$x^I$};
			\draw[-Latex] (0,0) -- (0,1.5) node[above] {$y^I$};
			\draw[dotted] (0,.5) -- (1,.5) ;
						  
			\draw (0,0) rectangle (1,1);
			\draw[thick, Red] (0,0) -- (1,0) node[midway, below] {$[a^I]$}; 
			\draw[thick, RoyalBlue] (0,0) -- (0,1) node[midway, left] {$[b^I]$}; 
						   
			\node at (.5,1.5) {$\beta^I = 0$}; 
			\node at (1,0) [below]{$R_1^I$};
			\node at (0,1) [left]{$R_2^I$};
			\fill[darkgreen] (0,0) circle (2pt) node[above right]{1};
			\fill[darkgreen] (0,.5) circle (2pt) node[above right]{2};
			\fill[darkgreen] (.5,0) circle (2pt) node[above]{3};
			\fill[darkgreen] (.5,.5) circle (2pt) node[above]{4};
		\end{scope}
		\begin{scope}[xshift=7cm]
			\draw[-Latex] (0,0) -- (1.5,0) node[right] {$x^I$};
			\draw[-Latex] (0,0) -- (0,1.5) node[above] {$y^I$};
			\draw[dotted] (0,.5) -- (1,.5) ;
			\draw[dotted] (0,1) -- (1,1) ;
						
			\draw (0,0) -- (1,.5) -- (1,1.5) -- (0,1);
			\draw[thick, Red] (0,0) -- (1,0) node[midway, below] {$[a^I]$}; 
			\draw[thick, RoyalBlue] (0,0) -- (0,1) node[midway, left] {$[b^I]$}; 
			\draw[thick, Green] (0,0) -- (1,.5) node[near end, below] {$[a'^I]$};
						  
			\node at (.5,1.5) {$\beta^I = 1$}; 
			\draw (1,-0.025) -- (1,0.025) ;
			\node at (1,0) [below]{$R_1^I$}; 
			\node at (0,1) [left]{$R_2^I$}; 
			\fill[darkgreen] (0,0) circle (2pt) node[above right, yshift=0.05cm]{1};
			\fill[darkgreen] (0,.5) circle (2pt) node[above right]{2};
			\fill[darkgreen] (.5,.25) circle (2pt) node[above]{3};
			\fill[darkgreen] (.5,.75) circle (2pt) node[above]{4};
		\end{scope}
	\end{tikzpicture}  
	\caption{The $\Z_2$ invariant \textbf{a}-type (left) and \textbf{b}-type (right) lattices.
		$\Z_2$ fixed points $\{1,2,3,4\}$ are shown as blobs.
		The $\R$ invariant $x^I$ axis is along the 1-cycle $[a^I] - \frac{\beta^I}{2} [b^I]$ with $\beta^I=0$ for the \textbf{a}-type lattice and $\beta^I=1$ for the \textbf{b}-type lattice.
		$\R$ acts as reflection along the $y^I$ axis, which is spanned by the 1-cycle $[b^I]$.
		For the \textbf{a}-type lattice, all $\Z_2$ fixed points are invariant under $\R$, whereas
		for the \textbf{b}-type lattice, only 1 and 2 are invariant while $3 \stackrel\R{\leftrightarrow} 4$.}
	\label{Fig:Z2-lattice}
\end{figure}

Let us determine how $\Omega \R$ acts on the various 3-cycles. For the untwisted cycles this is straightforward because the horizontally placed O-planes reflect the vertical axis only,
\begin{align}\label{untwR}
	\Omega \R:\begin{cases} 
	[a^I]\to [a^I] ~,       \\
	[b^I]\to -[b^I] ~.      
	\end{cases}             
\end{align}
Therefore, the wrapping numbers are mapped as $\Omega \R:(n^I_a,\widetilde m^I_a)\to (n^I_a,-\widetilde m^I_a)$, respectively $\Omega \R:(n^I_a,m^I_a)\to (n^I_a,-m^I_a-\beta^I\,n^I_a)$.

For the twisted sector 3-cycles, the canonical action  $\Omega \R$ corresponding to the models without vector structure including the signs $\eta_{\Omega \R}, \eta_{\Omega \R g}$ consistent with \eqref{opsigns} reads as,
\begin{align}\label{twR}
	\Omega \R:\begin{cases}                                                                   
	\alpha^g_{ij,\,n} \to -\eta_{\Omega \R}\, \eta_{\Omega \R g} \, \alpha^g_{\R(i)\R(j),\,n} \\
	\alpha^g_{ij,\,m} \to  \eta_{\Omega \R}\, \eta_{\Omega \R g}\, \alpha^g_{\R(i)\R(j),\,m}  
	\end{cases}                                                                               
\end{align}
where for $\beta^I=0$ the reflection $\R$ leaves all four fixed points $i\in\{1,2,3,4\}$ invariant, whereas for $\beta^I=1$ the action $\R$ interchanges the fixed points 3 and 4 while leaving 1 and 2 unchanged,
\begin{equation}\label{eq:R34}
	\R:  \begin{cases}
	1 \rightarrow 1, \\
	2 \rightarrow 2, \\
	3 \rightarrow 3+\beta^I ,\\
	4 \rightarrow 4-\beta^I ,
	\end{cases} \qquad \beta^I \in \{0,1\},
\end{equation}
as can be seen from the figure \ref{Fig:Z2-lattice}.

Implementing the $3\leftrightarrow 4$ interchange \eqref{eq:R34} coming from the $\Omega \R$ action on the number of common fixed points \eqref{eq:deltaij}, we get,
\begin{align}\label{eq:deltapij}
	\delta_{a b'}^g\equiv \delta^g(\alpha^{a}_{ij}, \alpha^{b'}_{kl}) & = \delta^g_{S^{a}_{i},S^{b'}_{k}} \delta^g_{S^{a}_{j},S^{b'}_{l}}. 
\end{align}
Again assuming that every fractional brane intersects the origin the above relation simplifies as,
\begin{align}\label{eq:deltap}
	\delta_{a b'}^g & = \sum _{i=1}^2 \sum _{j=1}^2 \sum _{k=1}^2 \sum _{l=1}^2 \delta^g_{S^{a}_{1,i},S^{b'}_{1,k}} \delta^g_{S^{a}_{2,j},S^{b'}_{2,l}}~. 
\end{align}

Using equations \eqref{untwR}, \eqref{twR} and \eqref{eq:deltapij} the intersection number of $\Pi_a^F$ with the $\Omega \R$ image of $\Pi_b^F$ can be computed as,
\begin{align} \label{Iabp}
	\Pi_a^F \circ \left(\Pi_b^F\right)' & = \eta_{\Omega \R}\,\frac{1}{4}\left(-\eta_{\Omega \R}\,2^{-k}\prod_I (n^I_a l_b^I + l_a^I n^I_b) +\!\!\! \sum _{g\in \{\theta, \omega,\theta\omega\}} \!\!\!\eta_{\Omega \R g}\, \delta_{ab'}^g \,2^{-\beta^{g}} (n^g_a\,l_b^g + l_a^g\,n^g_b) \right). 
\end{align}
As a special case of \eqref{eq:deltap}, setting $a=b$ implies,
\begin{align}\label{eq:gij}
	\delta_{aa'}^g & = \sum_{h<i} \delta_{a a}^g \frac{\left|\epsilon^{ghi}\right|\,Q^h_{a}\, Q^i_{a}}{4}, 
\end{align}
where $Q^I_{a}$ denotes the number of invariant fixed points on the $I^\text{th}$ $\T^2$ under the $\Omega \R$ given as,
\begin{align}\label{Q}
	Q^I_a & \equiv \frac{3}{2} + \frac{(-1)^{\beta^I n^I_a}}{2}, \qquad \beta^I \in \{0,1\}, 
\end{align}
which is equal to two, except for the case $\beta^I=1$ and $n_a^I=$ odd, where it is equal to one.

Setting $a=b$ in \eqref{Iabp} and using \eqref{eq:gij} the intersection number of $\Pi_a^F$ with its own $\Omega \R$ image is,
\begin{align} \label{Iaap}
\Pi_a^F\circ \left(\Pi_a^F\right)' & = \eta_{\Omega \R}\, \Bigl(-\eta_{\Omega \R}\,2^{1-k}\prod_I n^I_a\, l_a^I +\eta_{\Omega \R\theta}\, \frac{Q^1_{a}\, Q^2_{a}}{ 2}\,2^{-\beta^{3}} n_a^3\, l_a^3  \nonumber \\
&\quad\qquad~~ +\eta_{\Omega \R\omega}\, \frac{Q^2_{a}\, Q^3_{a}}{2}\,2^{-\beta^{1}} n_a^1\, l_a^1 +\eta_{\Omega \R\theta\omega}\, \frac{Q^1_{a}\, Q^3_{a}}{ 2}\,2^{-\beta^{2}} n_a^2\, l_a^2 \Bigr)\,.                                                                                                                                                
\end{align} 

For the intersection with the orientifold plane one obtains,
 
\begin{align} \label{IaO6}
\Pi_{a}\circ \Pi_\textrm{O6}^F &= 2^{1-k}\,\Bigl(-\eta_{\Omega \R}\,\prod_I l_a^I + \eta_{\Omega \R\theta}\,  n_a^1\, n_a^2\, l_a^3 + \eta_{\Omega \R\omega}\, l_a^1 \, n_a^2\, n_a^3 + \eta_{\Omega \R\theta\omega}\, n_a^1\, l_a^2 \, n_a^3 \Bigr). 
\end{align}

Using the expressions in Eqs.~\eqref{Iab}, \eqref{Iabp}, \eqref{Iaap}, and \eqref{IaO6}, one can determine the chiral spectrum involving symmetric and antisymmetric representations of the gauge group $\prod_a \U(N_a)$. In particular, a stack of $N_a$ fractional D6-branes satisfying the condition $\Omega \mathcal{R} \Pi_a^F = \Pi_a^F$ gives rise to a $\USp(2N_a)$ gauge symmetry.

In our construction of Pati-Salam models, we employ four visible sector rigid branes denoted by $a, b, c, d$, in contrast to the usual case which involves only three branes and does not include twisted sectors. Consequently, the three-family condition is modified and takes the following form:
\begin{equation}\label{eq:3families}
	I_{ab} + I_{ab'} = - \left( I_{ac} + I_{ac'} + I_{ad} + I_{ad'} \right) = \pm 3 \,,
\end{equation}
where positive intersection numbers, in our convention, refer to left-chiral supermultiplets.

\subsection{Supersymmetry conditions}\label{susyconstraints}

Let us define the following products of wrapping numbers,
\begin{alignat}{4}
A_a         & \equiv -n_a^1n_a^2n_a^3, \qquad & B_a         & \equiv n_a^1l_a^2l_a^3, \qquad & C_a         & \equiv l_a^1n_a^2l_a^3, \qquad & D_a         & \equiv l_a^1l_a^2n_a^3, \nonumber \\
\tilde{A}_a & \equiv -l_a^1l_a^2l_a^3,        & \tilde{B}_a & \equiv l_a^1n_a^2n_a^3,        & \tilde{C}_a & \equiv n_a^1l_a^2n_a^3,        & \tilde{D}_a & \equiv n_a^1n_a^2l_a^3. \label{variables} 
\end{alignat}
Preserving $\mathcal{N}=1$ supersymmetry in four dimensions after compactification from ten-dimensions restricts the rotation angle of any D6-brane with respect to the orientifold plane to be an element of $\SU(3)$, i.e.
\begin{equation}
	\theta^a_1 + \theta^a_2 + \theta^a_3 = 0 \mod 2\pi ,
\end{equation}
where $\theta_i$ is the angle between the D6-brane and orientifold-plane in the $i^\text{th}$ 2-torus and $\chi_i=R^2_i/R^1_i$ are the complex structure moduli for the $i^\text{th}$ 2-torus,
\begin{align}\label{theta}
	\tan \theta^a_j = \chi_j \frac{\widetilde{m}^a_j}{n^a_j} = \chi_j \frac{2^{- \beta_j}l^a_j}{n^a_j} \,. 
\end{align}
$\mathcal{N}=1$ supersymmetry conditions are given as,
\begin{eqnarray}
	x_A\tilde{A}_a+x_B\tilde{B}_a+x_C\tilde{C}_a+x_D\tilde{D}_a=0,\nonumber\\
	\frac{A_a}{x_A}+\frac{B_a}{x_B}+\frac{C_a}{x_C}+\frac{D_a}{x_D} < 0, \label{susyconditions}
\end{eqnarray}
where $x_A=\lambda,\; x_B=2^{\beta_2+\beta_3}\cdot\lambda /\chi_2\chi_3,\; x_C=2^{\beta_1+\beta_3}\cdot\lambda /\chi_1\chi_3,\; x_D=2^{\beta_1+\beta_2}\cdot\lambda /\chi_1\chi_2$.

\subsection{Tadpole cancellation}\label{tadpoleconstraints}

Since D6-branes and O6-orientifold planes are the sources of Ramond-Ramond charges they are constrained by the Gauss's law in compact space implying the sum of D-brane and cross-cap RR-charges must vanishes
\begin{eqnarray}\label{RRtadpole}
	\sum_a N_a [\Pi_a]+\sum_a N_a \left[\Pi_{a'}\right]-4[\Pi_\textrm{O6}]=0,
\end{eqnarray}
where the last terms arise from the O6-planes, which have $-4$ RR charges in D6-brane charge units. RR tadpole constraint is sufficient to cancel the $\SU(N_a)^3$ cubic non-Abelian anomaly while $\U(1)$ mixed gauge and gravitational anomaly or $[\SU(N_a)]^2 \U(1)$ gauge anomaly can be cancelled by the Green-Schwarz mechanism, mediated by untwisted RR fields \cite{Green:1984sg}.

\begin{table}[t]
	\centering\renewcommand{\arraystretch}{1.4}
	$\begin{array}{|c|c|c|}
		\hline
		\text{Orientifold Action} & \text{O6-Plane} & (n^1,l^1)\times (n^2,l^2)\times (n^3,l^3)                    \\
		\hline\hline
		\Omega \R                 & 1               & (2^{\beta_1},0)\times (2^{\beta_2},0)\times (2^{\beta_3},0)  \\
		\hline
		\Omega \R\omega           & 2               & (2^{\beta_1},0)\times (0,-2^{\beta_2})\times (0,2^{\beta_3}) \\
		\hline
		\Omega \R\theta\omega     & 3               & (0,-2^{\beta_1})\times (2^{\beta_2},0)\times (0,2^{\beta_3}) \\
		\hline
		\Omega \R\theta           & 4               & (0,-2^{\beta_1})\times (0,2^{\beta_2})\times (2^{\beta_3},0) \\
		\hline
	\end{array}$
	\caption{The wrapping numbers for four O6-planes.}
	\label{tab:orientifold}
\end{table}
 
The twisted and the untwisted tadpole cancellation conditions are given by
\begin{align}
	  & \sum_a N_a n_a^1 n_a^2 n_a^3 = 16 \,\eta_{\Omega \R}\,, \nonumber                                                                                      \\
	  & \sum_a N_a n_a^i \widetilde{m}_a^j \widetilde{m}_a^k = -2^{4-\beta^j-\beta^k} \eta_{\Omega \R i}\,,  \quad i\neq j\neq k \in \{1,2,3\}\,, \nonumber \\
	  & \sum_a N_a n_a^i \left(\epsilon_{a,kl}^i -\eta_{\Omega \R}\, \eta_{\Omega \R i}\, \epsilon_{a,\R(k)\R(l)}^i\right) = 0 \,, \nonumber                                \\
	  & \sum_a N_a \widetilde{m}_a^i \left(\epsilon_{a,kl}^i + \eta_{\Omega \R}\, \eta_{\Omega \R i}\, \epsilon_{a,\R(k)\R(l)}^i\right) = 0 \,,\label{twtadp}               
\end{align}
where $N_a$ denotes the number of D6-branes on stack $a$ and the sum is a sum over all stacks of D6-branes. $\R(k)=k$ in case of an untilted torus and $\R(\{1,2,3,4\})=\{1,2,4,3\}$ in the tilted case. The twisted charge $\epsilon_{a,ij}^{\omega}$ is non-zero if and only if $ij \in S_{\omega}^a$, i.e., if the brane $a$ passes through the fixed point $ij$ in the $\omega$-twisted sector, and so on. The orientifold projection acts on the wrapping numbers and twisted charges as follows,
\begin{eqnarray}
	\widetilde{m}^I &\rightarrow & -\widetilde{m}^I \,, \nonumber \\
	\epsilon_{kl}^i &\rightarrow & -\eta_{\Omega \R}\, \eta_{\Omega \R i}\, \epsilon^i_{\R(k)\R(l)}\, .
\end{eqnarray}

Cancellation of RR tadpoles requires introducing a number of orientifold planes also called ``filler branes'' that trivially satisfy the four-dimensional $\mathcal{N}=1$ supersymmetry conditions. The filler branes belong to the hidden sector USp group and carry the same wrapping numbers as one of the O6-planes as shown in table \ref{tab:orientifold}. USp group is hence referred with respect to the non-zero $A$, $B$, $C$ or $D$-type.

Orientifolds also have discrete D-brane RR charges classified by the $\mathbb{Z}_2$ K-theory groups, which are subtle and invisible by the ordinary homology \cite{Cascales:2003zp, Marchesano:2004yq, Marchesano:2004xz}, which should also be taken into account \cite{Uranga:2000xp}. The K-theory conditions are,
\begin{align}
	\sum_a \tilde{A}_a  = \sum_a  N_a  \tilde{B}_a = \sum_a  N_a  \tilde{C}_a = \sum_a  N_a \tilde{D}_a = 0 \textrm{~mod~}4. \nonumber\\ \label{K-charges}
\end{align}

\subsection{Gauge couplings from complex structure moduli}

Dynamical supersymmetry breaking in D6-brane models derived from Type IIA orientifolds has been explored in~\cite{Cvetic:2003yd}. The Kähler potential is given by
\begin{equation}\label{eq: Kaehler_potential}
    K = - \ln(S + \overline{S}) - \sum_{i=1}^3 \ln(U^i + \overline{U}^i) \, .
\end{equation}
The complex structure moduli $U^i$ can be extracted from the supersymmetry conditions, as shown in~\cite{Sabir:2022hko},
\begin{align}\label{U-moduli}
    U^i & = \frac{4i \chi^i + 2\beta_i^2 \chi_i^2}{4 + \beta_i^2 \chi_i^2}, \qquad \chi^i \equiv \frac{R_2^i}{R_1^i} \, .
\end{align}
These moduli, expressed in the string theory basis, can be mapped to the field theory basis using $\{s, u^i\}$ as follows~\cite{Lust:2004cx}:
\begin{equation}\label{eq:sugra-string-basis}
\begin{split}
    \mathrm{Re}(s)   &= \frac{e^{-\phi_4}}{2\pi} \frac{\sqrt{\mathrm{Im}(U^1) \, \mathrm{Im}(U^2) \, \mathrm{Im}(U^3)}}{\left|U^1 U^2 U^3\right|} , \\
    \mathrm{Re}(u^1) &= \frac{e^{-\phi_4}}{2\pi} \sqrt{\frac{\mathrm{Im}(U^1)}{\mathrm{Im}(U^2) \, \mathrm{Im}(U^3)}} \left| \frac{U^2 U^3}{U^1} \right| , \\
    \mathrm{Re}(u^2) &= \frac{e^{-\phi_4}}{2\pi} \sqrt{\frac{\mathrm{Im}(U^2)}{\mathrm{Im}(U^1) \, \mathrm{Im}(U^3)}} \left| \frac{U^1 U^3}{U^2} \right| , \\
    \mathrm{Re}(u^3) &= \frac{e^{-\phi_4}}{2\pi} \sqrt{\frac{\mathrm{Im}(U^3)}{\mathrm{Im}(U^1) \, \mathrm{Im}(U^2)}} \left| \frac{U^1 U^2}{U^3} \right| .
\end{split}
\end{equation}
The four-dimensional dilaton is related to the moduli via
\begin{equation}
    2\pi e^{\phi_4} = \left( \mathrm{Re}(s) \, \mathrm{Re}(u^1) \, \mathrm{Re}(u^2) \, \mathrm{Re}(u^3) \right)^{-1/4}.
\end{equation}

The holomorphic gauge kinetic function for a D6--brane stack $x$ wrapping a supersymmetric 3--cycle is~\cite{Blumenhagen:2006ci}:
\begin{align}\label{eq:gauge-kinetic-f}
    f_x & = \frac{1}{4 k_x} \left( n_x^1 n_x^2 n_x^3 \, s 
    - \frac{n_x^1 l_x^2 l_x^3 \, u^1}{2^{\beta_2 + \beta_3}} 
    - \frac{l_x^1 n_x^2 l_x^3 \, u^2}{2^{\beta_1 + \beta_3}} 
    - \frac{l_x^1 l_x^2 n_x^3 \, u^3}{2^{\beta_1 + \beta_2}} \right),
\end{align}
where $s$ and $u^i$ are the four--dimensional dilaton and complex structure moduli, respectively, and $k_x$ is the Kac--Moody level of $G_x$: $\kappa_x = 1$ for $\text{U}(N_x)$ and $\kappa_x = 2$ for $\text{USp}(2N_x)$ or $\text{SO}(2N_x)$ \cite{Ginsparg:1987ee, Hamada:2014eia}. With this convention, the gauge coupling is given by
\begin{equation}
    g_x^{-2} = \mathrm{Re}(f_x).
\end{equation}

When two gauge factors $G_c$ and $G_d$ are Higgsed to their diagonal subgroup $G_R$, canonical normalization of the gauge kinetic terms implies
\begin{equation}
    g_R^{-2} = g_c^{-2} + g_d^{-2}
              = \mathrm{Re}(f_c) + \mathrm{Re}(f_d).
\end{equation}
Therefore, the holomorphic gauge kinetic function for the diagonal is simply \cite{Dienes:1996yh}
\begin{equation}
    f_R = f_c + f_d \,.
\end{equation}   
No factor of $1/2$ appears here: the dependence on the Kac–Moody levels $k_x$ is already included in each $f_x$ through the $1/(4k_x)$ prefactor in its definition \eqref{eq:gauge-kinetic-f}. For $k_c = k_d = 1$ one finds $k_R = k_c + k_d = 2$ for the diagonal group, but this is automatically encoded in $f_R$ above.

At tree level, the gauge couplings satisfy
\begin{equation}
\begin{split}
    g_a^2 &= \frac{\mathrm{Re}(f_b)}{\mathrm{Re}(f_a)}\, g_b^2
           = \frac{\mathrm{Re}(f_c) + \mathrm{Re}(f_d)}{\mathrm{Re}(f_a)} \, g_R^2 
           = k_Y \left(\frac{5}{3}\right) g_Y^2
           = \gamma \,\pi e^{\phi_4} \,,
\end{split}
\end{equation}
where $g_a$, $g_b$, and $g_Y$ are the strong, weak, and hypercharge couplings, $g_R$ is the coupling of the diagonal $\SU(2)_R$, $k_Y$ is the effective Kac–Moody level of the canonically normalized hypercharge, and $\gamma$ is a model-dependent constant fixed by the internal moduli.

\section{Three-family Pati-Salam models from rigid branes}\label{sec:models}

To construct the three family Pati-Salam model using rigid, semi-rigid and non-rigi.d branes we follow the strategy outlined in \cite{Forste:2008ex}. In the $\Z_2 \times \Z_2^{\prime}$ orbifold, fractional branes invariant under $\Omega \R$ are those placed on top of an exotic O$6^{(+,+)}$ plane that is taken as O$_{\Omega\R}$ in our choice \eqref{opsigns}. The adjoint fields from $aa$ sector do not arise for the rigid branes. All of the three two-tori are taken to be rectangular.  

We begin by outlining the model building strategy. As previously discussed, obtaining an odd number of chiral families requires that some of the fixed point contributions, denoted by $\delta^{g}_{ab}$, differ from their maximal values, i.e., $\delta^{g}_{ab} \neq (4,4,4)$, where four is the maximum contribution in each entry. Consequently, in order to cancel twisted RR tadpoles at all fixed points, it becomes necessary to introduce additional D6-branes beyond those used in the simpler case where all fixed points are shared among branes, i.e., $\delta^{g}_{ab} = (4,4,4)$~\cite{Forste:2008ex}.

These additional branes must be introduced carefully so as not to give rise to exotic chiral matter. In general, they are assigned to the hidden sector. However, in order to realize massless GUT Higgs pairs in the open string spectrum, it will be necessary to recombine one of these extra stacks with the stack initially responsible for the $\SU(2)_R$ gauge symmetry of the Pati-Salam model. Additionally, we restrict our attention to models preserving $\mathcal{N}=1$ supersymmetry, which imposes further constraints on the brane wrapping numbers and leads to partial stabilization of the closed string moduli.

More specifically, we consider an initial configuration consisting of four rigid D6-branes, labeled $\{a, b, c, d\}$, which constitute the (initially) \textit{visible} sector. These branes share some, but not all, fixed points in various twisted sectors, which generically results in uncancelled twisted tadpoles. To cancel these contributions, an additional (initially) \textit{hidden} sector of branes must be introduced. To minimize the number of required hidden sector branes, the three stacks $\{b, c, d\}$ which generate the $\SU(2)_L \times \SU(2)_R$ gauge symmetry are chosen to share exactly the same set of fixed points, i.e., $\delta^{g}_{bc} = (4,4,4)$. Consequently, their contributions to twisted tadpoles are identical and can be arranged to cancel one another. This leaves only the twisted tadpoles sourced by the stack $a$.

To cancel these remaining twisted tadpoles, we introduce additional stacks $\{e_1, e_2\}$ such that the total twisted tadpole contributions from the combined set $\{a, b, c, d, e_1, e_2\}$ cancel completely. The \textit{semi-rigid} branes $\{e_1, e_2\}$ are constructed to have identical wrapping numbers and twisted charges with respect to one of the $\mathbb{Z}_2$ orbifold factors, ensuring that their mutual contributions cancel internally. 

Finally, the cancellation of untwisted tadpoles is achieved by adding suitable hidden sector branes that do not source additional twisted tadpoles. Branes of this type $\{f_1, f_2, f_3, f_4\}$ can recombine into bulk branes and thereby form \textit{non-rigid} stacks. In fact, the requirement of unbroken $\mathcal{N}=1$ supersymmetry restricts all hidden sector branes to be either semi-rigid or non-rigid.

The anomalies associated with the four global $\U(1)$ symmetries embedded in $\U(4)_C$, $\U(2)_L$, $\U(2)_{R_1}$, and $\U(2)_{R_2}$ are canceled via the Green-Schwarz mechanism. As a result, the corresponding $\U(1)$ gauge bosons acquire masses through linear $B \wedge F$ couplings. Consequently, the effective gauge symmetry is $\SU(4)_C \times \SU(2)_L \times \SU(2)_{R_1}\times\SU(2)_{R_2}$. When the fields $\Delta_i$ acquire vacuum expectation values (VEVs), the $\SU(2)_{R_1}\times\SU(2)_{R_2}$ symmetry is broken down to the diagonal subgroup $\SU(2)_R$. This leads to the Pati-Salam gauge symmetry, $\SU(4)_C \times \SU(2)_L \times \SU(2)_R$.

We obtain three generations of left-handed fermions $F^i_L$, and three generations of right-handed fermions arising from the combination of $F_R^i$ and $F_R^{\prime i}$ according to \eqref{eq:3families}. By assigning suitable VEVs to $F_R^{ci}$ and a linear combination of $F_R^i$, the Pati-Salam gauge symmetry can be broken down to the Standard Model gauge symmetry. Moreover, this symmetry breaking can preserve four-dimensional $\mathcal{N}=1$ supersymmetry, provided D-flatness and F-flatness conditions are satisfied.

The Standard Model fermion masses and mixing, as well as vector-like masses for $F_R^{\prime ci}$ and the two remaining linear combinations of $F_R^i$, can be generated through the following superpotential,
\begin{align}
	W \supset Y_{ijk} F_L^i F_R^j \Phi^k + Y_{ijk}^{\prime} F_R^{\prime ci} F_R^j \Delta^k . 
\end{align}
Furthermore, the $\SU(4)$ gauge groups in the hidden sector exhibit negative beta functions, which allows for supersymmetry breaking via gaugino condensation.
 
Following the above outlined model-building strategy, the resulting models are presented in appendix \ref{appA} in tables~\ref{model:r06}, \ref{model:r08}, \ref{model:r10}, \ref{model:r19}, \ref{model:r20}, \ref{model:r22}, \ref{model:r25}, \ref{model:r26}, \ref{model:r29}, \ref{model:r30} and their respective particle spectra are tabulated in tables~\ref{spec:r06}, \ref{spec:r08}, \ref{spec:r10}, \ref{spec:r19}, \ref{spec:r20}, \ref{spec:r22}, \ref{spec:r25}, \ref{spec:r26}, \ref{spec:r29} and \ref{spec:r30} respectively. 

Note that in tabulating the full spectrum of each model, we have implemented a specific deformation to eliminate additional massless states in the hidden sector. Specifically, we have displaced the hidden sector D6-branes $e, f, g, \ldots$ from the origin of the internal space $\mathbb{T}^2 \times \mathbb{T}^2 \times \mathbb{T}^2$ to alternative orbifold fixed points as discussed in table~\ref{fixed}. This displacement leads to the vanishing of certain intersection numbers between the hidden sector branes and the visible sector branes $a, b, c, d$, thereby removing unwanted massless matter. Alternatively, the same effect can be achieved by introducing non-trivial discrete Wilson lines~\cite{Blumenhagen:2005mu, Forste:2010gw}.

\begin{table}[t]
	\centering
	$
\begin{array}{|c|D{.}{\times}{11}|c|}
		\hline
		\text{Model~\hyperref[model:r06]{r06}} & \mc{\text{Quantum Numbers}}                   & \text{Fields}        \\
		\hline
		ab'                                    & 3. (\bar{4},\bar{2},1,1)                      & F_L^{c\, i}(Q_L,L_L) \\
		ac                                     & 1. (4,1,\bar{2},1)                            & F_R^{c\, i}(Q_R,L_R) \\
		ac'                                    & 3. (4,1,2,1)                                  & F_R^{c\, i}(Q_R,L_R) \\
		ad'                                    & 2. (4,1,1,2)                                  & F_R^{c'i}(Q_R,L_R)   \\
		ad                                     & 3. (\bar{4},1,1,2)                            & F_R^{'i}(Q_R,L_R)    \\
		b_{\overline{\yng(1,1)_{}}}            & 30. (1,\bar{1}_{\overline{\yng(1,1)_{}}},1,1) & S_L^i                \\
		c_{\overline{\yng(1,1)_{}}}            & 25. (1,1,\bar{1}_{\overline{\yng(1,1)_{}}},1) & S_R^i                \\
\hline
	\end{array}$
	\caption{Particle spectrum of Model~\hyperref[model:r06]{r06} with gauge symmetry $\SU(4)_{C}\times\SU(2)_{L}\times\SU(2)_{R_1}\times\SU(2)_{R_2}$.}
	\label{spec:r06}
\end{table}


\begin{table}[t]
	\centering
	$\begin{array}{|c|D{.}{\times}{13}|c|}
		\hline
		\text{Model~\hyperref[model:r08]{r08}} & \mc{\text{Quantum Numbers}}                       & \text{Fields}        \\
		\hline
		ab                                     & 3. (4,\bar{2},1,1,1,1)                            & F_L^i(Q_L,L_L)       \\
		ac                                     & 2. (4,1,\bar{2},1,1,1)                            & F_R^{c\, i}(Q_R,L_R) \\
		ad'                                    & 1. (4,1,1,2,1,1)                                  & F_R^{c'i}(Q_R,L_R)   \\
		ad                                     & 6. (\bar{4},1,1,2,1,1)                            & F_R^{'i}(Q_R,L_R)    \\
		b_{\overline{\yng(1,1)_{}}}            & 30. (1,\bar{1}_{\overline{\yng(1,1)_{}}},1,1,1,1) & S_L^i                \\
		c_{\yng(1,1)_{}}                       & 15. (1,1,1_{\yng(1,1)_{}},1,1,1)                  & S_R^i                \\
		d_{\yng(1,1)_{}}                       & 20. (1,1,1,1_{\yng(1,1)_{}},1,1)                  & S_R^i                \\
\hline
	\end{array}$
	\caption{Particle spectrum of Model~\hyperref[model:r08]{r08} with gauge symmetry $\SU(4)_{C}\times\SU(2)_{L}\times\SU(2)_{R_1}\times\SU(2)_{R_2}\times\SU(2)^2$.}
	\label{spec:r08}
\end{table}  

\begin{table}[t]
	\centering \footnotesize
	\begin{tabular}{c}$\begin{array}{|c|c|D{.}{\times}{12}|c|}
		\hline\text{Model~\hyperref[model:r08]{r08}} & \multicolumn{3}{c|}{\SU(4)_{C}\times\SU(2)_{L}\times\SU(2)_{R_1}\times\SU(2)_{R_2}\times\SU(2)^2} \\
		\hline 
		\text{Force} & \text{Intersection} & \mc{\text{Exotic Particle}}  & \text{Confined Spectrum} \\ 
		\hline\SU(2)_e         & be_1                & 2. (1,\bar{2},1,1,2,1)       & (1,\bar{2}^2,1,1,1,1)             \\
		                       & be_2                & 2. (1,\bar{2},1,1,1,2)       &                                   \\
		                       & be_1'               & 2. (1,\bar{2},1,1,\bar{2},1) &                                   \\
		                       & be_2'               & 2. (1,\bar{2},1,1,1,\bar{2}) &                                   \\
		\hline
		\end{array}$
	\end{tabular}
	\caption{The composite particle spectrum for model~\hyperref[spec:r08]{r08} formed due to the strong forces in hidden sector.} 
	\label{exotic:r08}
\end{table}


\begin{table}[t]
	\centering
	$\begin{array}{|c|D{.}{\times}{18}|c|}
		\hline
		\text{Model~\hyperref[model:r10]{r10}} & \mc{\text{Quantum Numbers}}                     & \text{Fields}     \\
		\hline
		ab                                     & 3. (\bar{4},2,1,1,1,1,1,1)                      & F_L^i(Q_L,L_L)    \\
		ad'                                    & 2. (\bar{4},1,1,\bar{2},1,1,1,1)                & F_R^{'i}(Q_R,L_R) \\
		ac                                     & 1. (\bar{4},1,2,1,1,1,1,1)                      & F_R^i(Q_R,L_R)    \\
		ac'                                    & 6. (4,1,2,1,1,1,1,1)                            & F_R^i(Q_R,L_R)    \\
		b_{\yng(1,1)_{}}                       & 6. (1,1_{\yng(1,1)_{}},1,1,1,1,1,1)             & S_L^i             \\
		c_{\yng(1,1)_{}}                       & 60. (1,1,1_{\yng(1,1)_{}},1,1,1,1,1)            & S_R^i             \\
		d_{\yng(1,1)_{}}                       & 9. (1,1,1,1_{\yng(1,1)_{}},1,1,1,1)             & S_R^i             \\
\hline
	\end{array}$
	\caption{Particle spectrum of model~\hyperref[model:r10]{r10} with gauge symmetry $\SU(4)_{C}\times\SU(2)_{L}\times\SU(2)_{R_1}\times\SU(2)_{R_2}\times\SU(2)^4$.}
	\label{spec:r10}
\end{table}

\begin{table}[t]
	\centering \footnotesize
	\begin{tabular}{c}$\begin{array}{|c|c|D{.}{\times}{16}|c|}
		\hline\multicolumn{1}{|c|}{\text{Model~\hyperref[model:r10]{r10}}} & \multicolumn{3}{c|}{\SU(4)_{C}\times\SU(2)_{L}\times\SU(2)_{R_1}\times\SU(2)_{R_2}\times\SU(2)^4} \\
		\hline 
		\text{Force} & \text{Intersection} & \mc{\text{Exotic Particle}} & \text{Confined Spectrum} \\ 
		\hline\SU(2)_e         & ce_1                & 2. (1,1,2,1,\bar{2},1,1,1)  & (1,1,2^2,1,1,1,1,1)               \\
		                       & ce_2                & 2. (1,1,2,1,1,\bar{2},1,1)  &                                   \\
		                       & ce_1'               & 2. (1,1,2,1,2,1,1,1)        &                                   \\
		                       & ce_2'               & 2. (1,1,2,1,1,2,1,1)        &                                   \\
		\hline\SU(2)_f         & bf_1'               & 2. (1,2,1,1,1,1,2,1)        & (1,2^2,1,1,1,1,1,1)               \\
		                       & bf_2'               & 2. (1,2,1,1,1,1,1,2)        & (1,1,2^2,1,1,1,1,1)               \\
		                       & cf_1                & 4. (1,1,2,1,1,1,\bar{2},1)  &                                   \\
		                       & cf_2                & 4. (1,1,2,1,1,1,1,\bar{2})  &                                   \\
		                       & cf_1'               & 2. (1,1,2,1,1,1,2,1)        &                                   \\
		                       & cf_2'               & 2. (1,1,2,1,1,1,1,2)        &                                   \\
		\hline
		\end{array}$
	\end{tabular}
	\caption{The composite particle spectrum for model~\hyperref[spec:r10]{r10} formed due to the strong forces in hidden sector.} 
	\label{exotic:r10}
\end{table}


\begin{table}[t]
	\centering
	$\begin{array}{|c|D{.}{\times}{28}|c|}
		\hline
		\text{Model~\hyperref[model:r19]{r19}} & \mc{\text{Quantum Numbers}}                                     & \text{Fields}      \\
		\hline
		ab                                     & 1. (4,\bar{2},1,1,1,1,1,1,1,1,1,1,1)                            & F_L^i(Q_L,L_L)     \\
		ab'                                    & 2. (4,2,1,1,1,1,1,1,1,1,1,1,1)                                  & F_L^i(Q_L,L_L)     \\
		ad                                     & 1. (4,1,1,\bar{2},1,1,1,1,1,1,1,1,1)                            & F_R^{c'i}(Q_R,L_R) \\
		ac'                                    & 4. (\bar{4},1,\bar{2},1,1,1,1,1,1,1,1,1,1)                      & F_R^i(Q_R,L_R)     \\
		b_{\overline{\yng(1,1)_{}}}            & 12. (1,\bar{1}_{\overline{\yng(1,1)_{}}},1,1,1,1,1,1,1,1,1,1,1) & S_L^i              \\
		c_{\overline{\yng(1,1)_{}}}            & 24. (1,1,\bar{1}_{\overline{\yng(1,1)_{}}},1,1,1,1,1,1,1,1,1,1) & S_R^i              \\
		d_{\overline{\yng(1,1)_{}}}            & 24. (1,1,1,\bar{1}_{\overline{\yng(1,1)_{}}},1,1,1,1,1,1,1,1,1) & S_R^i              \\
\hline
	\end{array}$
	\caption{Particle spectrum of Model~\hyperref[model:r19]{r19} with gauge symmetry $\SU(4)_{C}\times\SU(2)_{L}\times\SU(2)_{R_1}\times\SU(2)_{R_2}\times\SU(2)^5\times\USp(4)^4$.}
	\label{spec:r19}
\end{table}

\begin{table*}[t]
	\footnotesize\centering 
	%
	\caption{The composite particle spectrum for model~\hyperref[spec:r19]{r19} formed due to the strong forces in hidden sector.} 
	\label{exotic:r19}
\end{table*}


\begin{table}[t]
	\centering
	$\begin{array}{|c|D{.}{\times}{28}|c|}
		\hline
		\text{Model~\hyperref[model:r20]{r20}} & \mc{\text{Quantum Numbers}}                                       & \text{Fields}        \\
		\hline
		ab                                     & 2. (\bar{4},2,1,1,1,1,1,1,1,1,1,1,1,1)                            & F_L^{c\, i}(Q_L,L_L) \\
		ab'                                    & 1. (\bar{4},\bar{2},1,1,1,1,1,1,1,1,1,1,1,1)                      & F_L^{c\, i}(Q_L,L_L) \\
		ad'                                    & 4. (4,1,1,2,1,1,1,1,1,1,1,1,1,1)                                  & F_R^{c'i}(Q_R,L_R)   \\
		ac'                                    & 1. (\bar{4},1,\bar{2},1,1,1,1,1,1,1,1,1,1,1)                      & F_R^i(Q_R,L_R)       \\
		b_{\overline{\yng(1,1)_{}}}            & 8. (1,\bar{1}_{\overline{\yng(1,1)_{}}},1,1,1,1,1,1,1,1,1,1,1,1)  & S_L^i                \\
		c_{\overline{\yng(1,1)_{}}}            & 24. (1,1,\bar{1}_{\overline{\yng(1,1)_{}}},1,1,1,1,1,1,1,1,1,1,1) & S_R^i                \\
		d_{\yng(1,1)_{}}                       & 24. (1,1,1,1_{\yng(1,1)_{}},1,1,1,1,1,1,1,1,1,1)                  & S_R^i                \\
\hline
	\end{array}$
	\caption{Particle spectrum of Model~\hyperref[model:r20]{r20} with gauge symmetry $\SU(4)_{C}\times\SU(2)_{L}\times\SU(2)_{R_1}\times\SU(2)_{R_2}\times\SU(2)^6\times\USp(4)^4$.}
	\label{spec:r20}
\end{table}  
 
\begin{table*}[t]
	\centering \footnotesize
	%
	\caption{The composite particle spectrum for model~\hyperref[spec:r20]{r20} formed due to the strong forces in hidden sector.} 
	\label{exotic:r20}
\end{table*}


\begin{table}[t]
	\centering
	$\begin{array}{|c|D{.}{\times}{18}|c|}
		\hline
		\text{Model~\hyperref[model:r22]{r22}} & \mc{\text{Quantum Numbers}}                     & \text{Fields}      \\
		\hline
		ab'                                    & 3. (4,2,1,1,1,1,1,1)                            & F_L^i(Q_L,L_L)     \\
		ad'                                    & 3. (4,1,1,2,1,1,1,1)                            & F_R^{c'i}(Q_R,L_R) \\
		ad                                     & 2. (\bar{4},1,1,2,1,1,1,1)                      & F_R^{'i}(Q_R,L_R)  \\
		ac                                     & 3. (\bar{4},1,2,1,1,1,1,1)                      & F_R^i(Q_R,L_R)     \\
		ac'                                    & 1. (\bar{4},1,\bar{2},1,1,1,1,1)                & F_R^i(Q_R,L_R)     \\
		b_{\yng(1,1)_{}}                       & 4. (1,1_{\yng(1,1)_{}},1,1,1,1,1,1)             & S_L^i              \\
		c_{\yng(1,1)_{}}                       & 15. (1,1,1_{\yng(1,1)_{}},1,1,1,1,1)            & S_R^i              \\
\hline
	\end{array}$
	\caption{Particle spectrum of Model~\hyperref[model:r22]{r22} with gauge symmetry $\SU(4)_{C}\times\SU(2)_{L}\times\SU(2)_{R_1}\times\SU(2)_{R_2}\times\USp(8)^4$.}
	\label{spec:r22}
\end{table}

\begin{table}[t]
	\centering \footnotesize
	\begin{tabular}{c}$\begin{array}{|c|c|D{.}{\times}{16}|c|}
		\hline \multicolumn{1}{|c|}{\text{Model~\hyperref[model:r22]{r22}}} & \multicolumn{3}{c|}{\SU(4)_{C}\times\SU(2)_{L}\times\SU(2)_{R_1}\times\SU(2)_{R_2}\times\USp(8)^4} \\
		\hline 
		\text{Force} & \text{Intersection} & \mc{\text{Exotic Particle}}      & \text{Confined Spectrum} \\ 
		\hline
		\USp(8)_e              & ce_1                & 1. (1,1,\bar{2},1,8,1,1,1)       & (1,1,\bar{2}^2,1,1,1,1,1)         \\
		                       & ce_3                & 1. (1,1,\bar{2},1,1,1,8,1)       & (1,1,1,\bar{2}^2,1,1,1,1)         \\
		                       & ce_1'               & 1. (1,1,\bar{2},1,\bar{8},1,1,1) &                                   \\ 
		                       & ce_3'               & 1. (1,1,\bar{2},1,1,1,\bar{8},1) &                                   \\
		                       & de_4                & 1. (1,1,1,\bar{2},1,1,1,8)       &                                   \\
		                       & de_4'               & 1. (1,1,1,\bar{2},1,1,1,\bar{8}) &                                   \\
		\hline
		\end{array}$
	\end{tabular}
	\caption{The composite particle spectrum for model~\hyperref[spec:r22]{r22} formed due to the strong forces in hidden sector.} 
	\label{exotic:r22}
\end{table}


\begin{table}[t]
	\centering
	$\begin{array}{|c|D{.}{\times}{23}|c|}
		\hline
		\text{Model~\hyperref[model:r25]{r25}} & \mc{\text{Quantum Numbers}}                                 & \text{Fields}     \\
		\hline
		ab'                                    & 3. (4,2,1,1,1,1,1,1,1,1,1)                                  & F_L^i(Q_L,L_L)    \\
		ad                                     & 1. (\bar{4},1,1,2,1,1,1,1,1,1,1)                            & F_R^{'i}(Q_R,L_R) \\
		ac'                                    & 2. (\bar{4},1,\bar{2},1,1,1,1,1,1,1,1)                      & F_R^i(Q_R,L_R)    \\
		b_{\overline{\yng(1,1)_{}}}            & 10. (1,\bar{1}_{\overline{\yng(1,1)_{}}},1,1,1,1,1,1,1,1,1) & S_L^i             \\
		c_{\overline{\yng(1,1)_{}}}            & 40. (1,1,\bar{1}_{\overline{\yng(1,1)_{}}},1,1,1,1,1,1,1,1) & S_R^i             \\
		d_{\yng(1,1)_{}}                       & 40. (1,1,1,1_{\yng(1,1)_{}},1,1,1,1,1,1,1)                  & S_R^i             \\
		cd                                     & 24. (1,1,\bar{2},2,1,1,1,1,1,1,1)                           & \Delta ^i         \\
\hline
	\end{array}$
	\caption{Particle spectrum of Model~\hyperref[model:r25]{r25} with gauge symmetry $\SU(4)_{C}\times\SU(2)_{L}\times\SU(2)_{R_1}\times\SU(2)_{R_2}\times\SU(2)^3\times\USp(8)^4$.}
	\label{spec:r25}
\end{table}

\begin{table*}[t]
	\centering \footnotesize
	%
	\caption{The composite particle spectrum of model~\hyperref[spec:r25]{r25} formed due to the strong forces in hidden sector.} 
	\label{exotic:r25}
\end{table*}


\begin{table}[t]
	\centering
	$\begin{array}{|c|D{.}{\times}{24}|c|}
		\hline
		\text{Model~\hyperref[model:r26]{r26}} & \mc{\text{Quantum Numbers}}                                 & \text{Fields}        \\
		\hline
		ab'                                    & 1. (\bar{4},\bar{2},1,1,1,1,1,1,1,1,1)                      & F_L^{c\, i}(Q_L,L_L) \\
		ab                                     & 4. (4,\bar{2},1,1,1,1,1,1,1,1,1)                            & F_L^i(Q_L,L_L)       \\
		ad                                     & 1. (\bar{4},1,1,2,1,1,1,1,1,1,1)                            & F_R^{'i}(Q_R,L_R)    \\
		ac                                     & 2. (\bar{4},1,2,1,1,1,1,1,1,1,1)                            & F_R^i(Q_R,L_R)       \\
		b_{\overline{\yng(1,1)_{}}}            & 48. (1,\bar{1}_{\overline{\yng(1,1)_{}}},1,1,1,1,1,1,1,1,1) & S_L^i                \\
		c_{\yng(1,1)_{}}                       & 2. (1,1,1_{\yng(1,1)_{}},1,1,1,1,1,1,1,1)                   & S_R^i                \\
		d_{\overline{\yng(1,1)_{}}}            & 4. (1,1,1,\bar{1}_{\overline{\yng(1,1)_{}}},1,1,1,1,1,1,1)  & S_R^i                \\
\hline
	\end{array}$
	\caption{Particle spectrum of Model~\hyperref[model:r26]{r26} with gauge symmetry $\SU(4)_{C}\times\SU(2)_{L}\times\SU(2)_{R_1}\times\SU(2)_{R_2}\times\USp(4)\times\SU(4)^6$.}
	\label{spec:r26}
\end{table}

\begin{table*}[t]
	\centering \footnotesize
	%
	\caption{The composite particle spectrum of model~\hyperref[spec:r26]{r26} formed due to the strong forces in hidden sector.} 
	\label{exotic:r26}
\end{table*}


\begin{table}[t]
	\centering
	$\begin{array}{|c|D{.}{\times}{31}|c|}
		\hline
		\text{Model~\hyperref[model:r29]{r29}} & \mc{\text{Quantum Numbers}}                                         & \text{Fields}        \\
		\hline
		ab                                     & 3. (4,\bar{2},1,1,1,1,1,1,1,1,1,1,1,1,1)                            & F_L^i(Q_L,L_L)       \\
		ac                                     & 1. (4,1,\bar{2},1,1,1,1,1,1,1,1,1,1,1,1)                            & F_R^{c\, i}(Q_R,L_R) \\
		ad                                     & 2. (4,1,1,\bar{2},1,1,1,1,1,1,1,1,1,1,1)                            & F_R^{c'i}(Q_R,L_R)   \\
		ac'                                    & 6. (\bar{4},1,\bar{2},1,1,1,1,1,1,1,1,1,1,1,1)                      & F_R^i(Q_R,L_R)       \\
		b_{\overline{\yng(1,1)_{}}}            & 4. (1,\bar{1}_{\overline{\yng(1,1)_{}}},1,1,1,1,1,1,1,1,1,1,1,1,1)  & S_L^i                \\
		c_{\overline{\yng(1,1)_{}}}            & 60. (1,1,\bar{1}_{\overline{\yng(1,1)_{}}},1,1,1,1,1,1,1,1,1,1,1,1) & S_R^i                \\
		c_{\overline{\yng(2)}}                 & 6. (1,1,\bar{3}_{\overline{\yng(2)}},1,1,1,1,1,1,1,1,1,1,1,1)       & T_R^i                \\
\hline
	\end{array}$
	\caption{Particle spectrum of Model~\hyperref[model:r29]{r29} with gauge symmetry $\SU(4)_{C}\times\SU(2)_{L}\times\SU(2)_{R_1}\times\SU(2)_{R_2}\times\SU(2)^3\times\SU(4)^4\times\USp(4)^4$.}
	\label{spec:r29}
\end{table}

\begin{table*}[t]
	\centering \footnotesize
	%
	\caption{The composite particle spectrum of model~\hyperref[spec:r29]{r29} formed due to the strong forces in hidden sector.} 
	\label{exotic:r29}
\end{table*}


\begin{table}[t]
	\centering
	$\begin{array}{|c|D{.}{\times}{18}|c|}
		\hline
		\text{Model~\hyperref[model:r30]{r30}} & \mc{\text{Quantum Numbers}}                           & \text{Fields}     \\
		\hline
		ab                                     & 3. (4,\bar{2},1,1,1,1,1,1)                            & F_L^i(Q_L,L_L)    \\
		ad                                     & 2. (\bar{4},1,1,2,1,1,1,1)                            & F_R^{'i}(Q_R,L_R) \\
		ad'                                    & 3. (4,1,1,2,1,1,1,1)                                  & F_R^{'i}(Q_R,L_R) \\
		ac                                     & 1. (\bar{4},1,2,1,1,1,1,1)                            & F_R^i(Q_R,L_R)    \\
		ac'                                    & 3. (\bar{4},1,\bar{2},1,1,1,1,1)                      & F_R^i(Q_R,L_R)    \\
		b_{\overline{\yng(1,1)_{}}}            & 30. (1,\bar{1}_{\overline{\yng(1,1)_{}}},1,1,1,1,1,1) & S_L^i             \\
		c_{\yng(1,1)_{}}                       & 20. (1,1,1_{\yng(1,1)_{}},1,1,1,1,1)                  & S_R^i             \\
\hline
	\end{array}$
	\caption{Particle spectrum of model~\hyperref[model:r30]{r30} with gauge symmetry $\SU(4)_{C}\times\SU(2)_{L}\times\SU(2)_{R_1}\times\SU(2)_{R_2}\times\USp(12)^4$.}
	\label{spec:r30}
\end{table}

\begin{table}[t]
	\centering \footnotesize
	\begin{tabular}{c}$\begin{array}{|c|c|D{.}{\times}{17}|c|}
		\hline \multicolumn{1}{|c|}{\text{Model~\hyperref[model:r30]{r30}}} & \multicolumn{3}{c|}{\SU(4)_{C}\times\SU(2)_{L}\times\SU(2)_{R_1}\times\SU(2)_{R_2}\times\USp(12)^4} \\
		\hline 
		\text{Force} & \text{Sector} & \mc{\text{Exotic Particle}}            & \text{Confined Spectrum} \\
		\hline\USp(12)_e       & be_3                & 1. (1,2,1,1,1,1,\overline{12},1)       & (1,2^2,1,1,1,1,1,1)               \\
		                       & be_3'               & 1. (1,2,1,1,1,1,12,1)                  & (1,1,\bar{2}^2,1,1,1,1,1)         \\ 
		                       & ce_2                & 1. (1,1,\bar{2},1,1,12,1,1)            & (1,1,1,\bar{2}^2,1,1,1,1)         \\ 
		                       & ce_2'               & 1. (1,1,\bar{2},1,1,\overline{12},1,1) &                                   \\ 
		                       & de_1                & 1. (1,1,1,\bar{2},12,1,1,1)            &                                   \\ 
		                       & de_1'               & 1. (1,1,1,\bar{2},\overline{12},1,1,1) &                                   \\\hline
		\end{array}$
	\end{tabular}
	\caption{The composite particle spectrum of model~\hyperref[spec:r30]{r30} formed due to the strong forces in hidden sector.} 
	\label{exotic:r30}
\end{table} 

\subsection{Model~\hyperref[spec:r06]{r06}}
  
The first model presented in table~\ref{model:r06} is a gauge theory of rank 6, involving a set of four rigid D6-branes labeled $\{a, b, c, d\}$, which share the same fixed points, i.e., $\delta_{ab}^g \neq (4, 4, 4)$. The model has a gauge symmetry $\SU(4)_C \times \SU(2)_L \times \SU(2)_{R_1} \times \SU(2)_{R_2}$. The particle spectrum, listed in table~\ref{spec:r06}, categorizes the various matter fields according to their quantum numbers under the gauge symmetry, including left- and right-handed fermions, scalar fields, and Higgs-like particles. Each field is labeled based on its origin from different brane intersections, such as $ab$, $ac$, $ad'$, etc., corresponding to chiral fermions in the fundamental representations of the gauge factors. The tree-level string scale gauge coupling relation and the complex structure moduli of the three two-tori are provided in the caption of table~\ref{model:r06}.

The particle spectrum of model~\hyperref[model:r06]{r06} is presented in table~\ref{spec:r06}. The table lists the quantum numbers of the matter fields under this gauge symmetry along with their corresponding field representations. The left-handed fermions, denoted as $F_L^i(Q_L, L_L)$, transform as $(\bar{4}, \bar{2},1,1)$, while the right-handed fermions appear in multiple representations: $F_R^{'i}(Q_R, L_R)$ under $(4,1,1,\bar{2})$ and $(4,1,1,2)$, as well as $F_R^i(Q_R, L_R)$ under $(\bar{4},1,2,1)$ and $(4,1,2,1)$. 

Additionally, singlet fields such as $S_L^i$ and $S_R^i$ are included, which transform under the antisymmetric representations of $\SU(2)_L$, $\SU(2)_{R_1}$, and $\SU(2)_{R_2}$. The Higgs-like fields, $\Phi^i(H_u, H_d)$, appear in the representations $(1, \bar{2}, 2, 1)$ and $(1, \bar{2}, \bar{2}, 1)$. Furthermore, scalar fields $\Delta^i$ serve as GUT Higgs fields responsible for breaking the Pati–Salam symmetry to the Standard Model gauge group. Besides, there exist several exotic particle states $\Xi ^{i}$ with quantum numbers $(1,2,1,\bar{2})$ and $(1,\bar{2},1,\bar{2})$. 

\subsection{Model~\hyperref[spec:r08]{r08}}

The model presented in table~\ref{model:r08} consists of a set of four rigid D6-branes $\{a, b, c, d\}$, which share same fixed points i.e $\delta_{ab}^g \neq (4, 4, 4)$. However, this leads to the uncancelled twisted tadpoles among themselves. We introduce two semi-rigid branes $e_1$ and $e_2$ related by $\Z_2$ image from $\theta$-twist such that all the twisted and untwisted charges vanish. Table~\ref{spec:r08} presents the particle spectrum of model~\hyperref[model:r08]{r08}, which extends the previous gauge symmetry by an additional $\SU(2)^2$, leading to the group $\SU(4)_{C} \times \SU(2)_{L} \times \SU(2)_{R_1} \times \SU(2)_{R_2} \times \SU(2)^2$. 

The left-handed fermions $F_L^i(Q_L, L_L)$ transform as $(4, \bar{2},1,1,1,1)$, while the right-handed fermions include multiple representations such as $F_R^{'i}(Q_R, L_R)$ in $(4,1,1,\bar{2},1,1)$ and $F_R^i(Q_R, L_R)$ in $(\bar{4},1,2,1,1,1)$ and $(4,1,2,1,1,1)$. The scalar sector contains multiple Higgs-like fields $\Phi^i(H_u,H_d)$ and $\Delta^i$ responsible for gauge symmetry breaking. Several exotic fields $X_L^i$ appear in the representations $(1,\bar{2},1,1,2,1)$ and $(1,\bar{2},1,1,1,2)$.

Table~\ref{exotic:r08} describes the composite particle spectrum in model~\hyperref[spec:r08]{r08}, arising from strong interactions in the hidden sector. The presence of the confining force $\SU(2)_e$ at the intersections $be_1$, $be_2$, $be_1'$, and $be_2'$ leads to the formation of composite states. As a result, all exotic particle states $X_L^i$ undergo confinement, giving rise to new bound states $(1, \bar{2}^2, 1, 1, 1, 1)$, which decouple from the low-energy spectrum.

\subsection{Model~\hyperref[spec:r10]{r10}}

For model~\hyperref[model:r10]{r10}, table~\ref{spec:r10} provides the detailed matter content under the gauge symmetry $\SU(4)_{C} \times \SU(2)_{L} \times \SU(2)_{R_1} \times \SU(2)_{R_2} \times \SU(2)^4$. This model features a rich particle spectrum, including multiple generations of fundamental fermions, singlet fields, and higher-dimensional representations. Additionally, several new exotic states $X_L^{i}$ and $X_R^{i}$ are introduced, which transform under extra hidden $\SU(2)$ sectors.

Table~\ref{exotic:r10} outlines the composite spectrum for model~\hyperref[model:r10]{r10}, describing the effects of confinement in the hidden sector. Exotic particles appearing at intersections such as $ce_1$ and $ce_2$ are subject to confinement under $\SU(2)_e$, leading to the formation of bound states that are neutral under the hidden gauge group. Similarly, intersections involving $bf_1'$ and $cf_1$ interact under the hidden $\SU(2)_f$ dynamics, resulting in additional confined composite states. The strong coupling dynamics thus causes the exotic particles to decouple from the low-energy spectrum.

\subsection{Model~\hyperref[spec:r19]{r19}}

The particle spectrum of model~\hyperref[model:r19]{r19} is presented in table~\ref{spec:r19}, which details the quantum numbers and associated fields under the gauge symmetry $\SU(4)_{C} \times \SU(2)_{L} \times \SU(2)_{R_1} \times \SU(2)_{R_2} \times \SU(2)^5 \times \USp(4)^4$.  

Table~\ref{exotic:r19} outlines the composite particle spectrum that emerges due to strong dynamics in the hidden sector. The exotic particles $X_L^{i}$ and $X_R^{i}$ undergo confinement under the non-Abelian gauge forces viz. $\SU(2)_e$, $\SU(2)_f$, $\SU(2)_g$, and $\USp(4)_h$. Each row specifies the initial exotic representation under the gauge symmetry and the confined particle spectrum resulting from the strong dynamics. 

\subsection{Model~\hyperref[spec:r20]{r20}} 

The particle spectrum of model~\hyperref[model:r20]{r20} is presented in table~\ref{spec:r20}, which lists the various chiral multiplets and their associated quantum numbers under the gauge symmetry $\SU(4)_{C}\times\SU(2)_{L}\times\SU(2)_{R_1}\times\SU(2)_{R_2}\times\SU(2)^6\times\USp(4)^4$.  

Table~\ref{exotic:r20} provides an overview of the composite particle spectrum that emerges due to confinement in the hidden sector. Several $\SU(2)$ and $\USp(4)$ gauge groups become strongly coupled, leading to the formation of bound states. For each confining force, $\SU(2)_e$, $\SU(2)_f$, and $\SU(2)_g$ and $\USp(4)_h$ the table specifies the exotic particle content involved in the strong interaction and the corresponding confined spectrum.  

\subsection{Model~\hyperref[spec:r22]{r22}}

The particle spectrum for model~\hyperref[model:r22]{r22} is presented in table~\ref{spec:r22}. The gauge symmetry of this model is given by $\SU(4)_{C} \times \SU(2)_{L} \times \SU(2)_{R_1} \times \SU(2)_{R_2} \times \USp(8)^4$. The table lists the various chiral matter fields, along with their corresponding quantum numbers under the gauge group.  

Table~\ref{exotic:r22} details the composite particle spectrum arising due to the strong dynamics in the hidden sector. Specifically, the $\USp(8)_e$ gauge interactions confine certain exotic particles, leading to the formation of bound states.

\subsection{Model~\hyperref[spec:r25]{r25}} 

Table~\ref{spec:r25} presents the chiral spectrum of model~\hyperref[model:r25]{r25}, characterized by the gauge symmetry $\SU(4)_{C} \times \SU(2)_{L} \times \SU(2)_{R_1} \times \SU(2)_{R_2} \times \SU(2)^3 \times \USp(8)^4$.  

Table~\ref{exotic:r25} details the confined composite states that emerge due to strong gauge dynamics in the hidden sector of model~\hyperref[model:r25]{r25}. These bound states result from the confining forces associated with $\SU(2)_e$, $\SU(2)_f$, and $\USp(8)_g$. 

\subsection{Model~\hyperref[spec:r26]{r26}}

The particle spectrum of model~\hyperref[model:r26]{r26} with gauge symmetry $\SU(4)_{C}\times\SU(2)_{L}\times\SU(2)_{R_1}\times\SU(2)_{R_2}\times\USp(4)\times\SU(4)^6$ is presented in table~\ref{spec:r26}. The model contains a variety of fields with their corresponding quantum numbers. These fields include left-handed fermions ($F_L^i$) and right-handed fermions ($F_R^i$), with the quantum numbers specified for each field under the different gauge groups. Additionally, there are scalar fields like $S_L^i$ and $S_R^i$, as well as exotic fields such as $\Delta^i$, $\Phi^i(H_u,H_d)$, and $\Xi^i$. The table also lists other types of fields, such as $X_L^{i}$, $X_R^{i}$, and various combinations of gauge group components associated with the exotic particles.

Table~\ref{exotic:r26} describes the composite particle spectrum formed in model~\hyperref[model:r26]{r26} due to the strong forces in the hidden sector. The exotic particles arise from different confining forces, each associated with specific intersections. For example, under the $\USp(4)_f$ force, composite particles like $(1, \bar{2}^2, 1, 1, 1, 1, 1, 1, 1, 1)$ are formed at the intersection of $bf_1$ and $bf_2$. Similarly, the $\SU(4)_g$ and $\SU(4)_h$ forces also contribute to the formation of exotic particles, including combinations of different gauge group components. The confined particle spectra are provided for each intersection, detailing the specific fields that combine to form the composite states.  

\subsection{Model~\hyperref[spec:r29]{r29}}

Table~\ref{spec:r29} presents the particle spectrum of model~\hyperref[model:r29]{r29} with the gauge symmetry $\SU(4)_{C} \times \SU(2)_{L} \times \SU(2)_{R_1} \times \SU(2)_{R_2} \times \SU(2)^3 \times \SU(4)^4 \times \USp(4)^4$.  

Table~\ref{exotic:r29} describes the composite particle spectrum of model~\hyperref[spec:r29]{r29} formed due to the strong forces in the hidden sector viz. $\SU(2)_e$, $\SU(2)_f$, $\SU(4)_g$, $\SU(4)_h$, and $\USp(4)_i$. For each confining force, the table lists the intersection types, the corresponding exotic particles formed, and the spectrum of confined particles.  
 
\subsection{Model~\hyperref[spec:r30]{r30}}

The particle spectrum of model~\hyperref[model:r30]{r30} is presented in table~\ref{spec:r30}. This model features a gauge symmetry group $\SU(4)_{C}\times\SU(2)_{L}\times\SU(2)_{R_1}\times\SU(2)_{R_2}\times\USp(12)^4$.  

Table~\ref{exotic:r30} provides the composite particle spectrum of model~\hyperref[spec:r30]{r30}, which arises due to the strong forces in the hidden sector of the model. The table describes the composite particles formed through the confining force $\USp(12)_e$.


\section{Asymptotic freedom}\label{sec:asymptoticfreedom}

Constructing the Standard Model from rigid cycles, in addition to eliminating adjoint chiral multiplets, is also advantageous for realizing a gauge theory that is asymptotically free, characterized by a negative one-loop beta function. This setup facilitates the convergence of the gauge couplings in the MSSM and also enables gaugino condensation through the non-perturbative superpotential of the form
\begin{align}
	W_{\text{eff}} & \sim \frac{M_\text{S}\,\beta_1^g}{32 \pi^2} \exp\left( \frac{8\pi^2}{g_\text{YM}^2\, \beta_1^g} \right), 
\end{align}
where the gauge couplings depend on the complex structure (or Kähler) moduli in Type IIA (or Type IIB) string theory. This effective superpotential may, in principle, stabilize some of the closed-string moduli, potentially in combination with other mechanisms such as background fluxes. 

In general, the beta functions are sensitive to the entire massless spectrum, including additional light non-chiral states that are not captured by intersection numbers or topological invariants. Therefore, it is crucial to have complete control over the full spectrum of the theory. To this end, let us examine the light spectrum arising from fractional branes, which are either constructed by splitting the bulk branes or are otherwise generic rigid branes.

\subsection{Fractional branes from splitting bulk branes}

Consider a bulk D-brane $a$ supporting a $\U(N)$ gauge group. This brane contains three adjoint chiral multiplets, in addition to other matter from intersections with other D-branes. Neglecting the latter, the one-loop beta function is
\begin{align}
	b_1^{\U(N)} = -3N + 3 \times N = 0, 
\end{align}
indicating that bulk D-branes have vanishing or positive beta functions.

One might attempt to improve this by splitting the bulk brane into four rigid fractional constituents ${b_1, b_2, b_3, b_4}$ transforming in the regular representation of $\mathbb{Z}_2 \times \mathbb{Z}_2$, forming, for example, the D-brane stack $b$. This decomposition yields a gauge group $\U(N)^4$ with no massless adjoint fields. However, additional nonchiral matter may arise between pairs of these fractional D-branes, and in fact, this is generally the case. This spectrum can be computed from the boundary state overlaps:
\begin{align} \label{eq:overlap}
	\tilde{A}_{b_i,b_j} &= \int_0^\infty dl\, \langle b_i| e^{-2\pi l H_{\text{cl}}} |b_j \rangle + \int_0^\infty dl\, \langle b_j| e^{-2\pi l H_{\text{cl}}} |b_i \rangle, \nonumber\\ 
&\qquad  \qquad \qquad   i \neq j,
\end{align}
whose loop channel representation is
\begin{equation}
	A_{b_i,b_j} = \int_0^\infty \frac{dt}{t} \, \text{Tr}_{ij+ji} \left( \frac{1 + \theta + \omega + \theta \omega}{4} \, e^{-2\pi t H_0} \right).
\end{equation}
The twisted sector projections result in a single massless hypermultiplet, composed of scalar states associated with the oscillators $\psi^I_{-\frac{1}{2}}|0\rangle$, $\overline{\psi}^I_{-\frac{1}{2}}|0\rangle$, for $I \in \{1,2,3\}$. For $i = j$, the non-compact oscillators $\psi^\mu_{-\frac{1}{2}}|0\rangle$, $\overline{\psi}^\mu_{-\frac{1}{2}}|0\rangle$ survive the projection, leading to an $\mathcal{N}=1$ vector multiplet.

Each $\U(N)$ factor thereby receives $6N$ chiral supermultiplets in the fundamental representation, resulting in a one-loop beta function:
\begin{align}
	b_1^{\U(N)} = -3N + 6N \times \frac{1}{2} = 0. 
\end{align}
Hence, splitting a $\U(N)$ bulk brane into fractional branes does not improve the asymptotic behavior of the beta function. Consequently, rigid D-branes obtained through this method cannot yield asymptotically free gauge groups.

\subsection{Generic rigid fractional branes}

Let us now consider a rigid D-brane $a$ that supports a $\U(4)$ gauge group and investigate whether the corresponding $\SU(4)$ subgroup is asymptotically free. Due to the rigidity of the brane, no adjoint fields appear in the non-chiral spectrum. However, from the $aa'$ sector, one hypermultiplet (i.e., two chiral multiplets) in the antisymmetric representation of $\SU(4)$ arises~\cite{Blumenhagen:2005tn}.

Potential extra massless states charged under $\SU(4)$ may arise from bifundamental vector-like pairs in the $ab$ sectors, where $b$ is another fractional D-brane. These states must be computed by accounting for twisted sector projections in the loop channel. Effectively, this involves computing the spectrum in the unorbifolded theory and then applying the orbifold projection. 

The resulting formula for the $\SU(4)$ beta function is given by
\begin{align}\label{eq:betaSU4}
	\beta^{\SU(4)} & = -3 \times 4 + N^\text{chiral}_a \times \frac{1}{2} + 2 \times 1 \,, 
\end{align}
where $N^\text{chiral}_a$ is twice the number of chiral multiplets arising from the intersection of the $\SU(4)$ brane with other visible sector branes:
\begin{align}
	N^\text{chiral}_a & = 2 \left(|I_{ab}| + |I_{ab'}| + |I_{ac}| + |I_{ac'}| + |I_{ad'}| + |I_{ad'}| \right)\,. 
\end{align}

Applying the relation \eqref{eq:betaSU4} to our set of models, we obtain the following beta function coefficients for the $\SU(4)$ gauge group:
\begin{align}
	\text{Model~\hyperref[spec:r06]{r06}}: \quad \beta^{\SU(4)} & = -3 \times 4 + 24 \times \frac{1}{2} + 2 \times 1 = 2 \,, \nonumber  \\
	\text{Model~\hyperref[spec:r08]{r08}}: \quad \beta^{\SU(4)} & = -3 \times 4 + 24 \times \frac{1}{2} + 2 \times 1 = 2 \,, \nonumber  \\
	\text{Model~\hyperref[spec:r10]{r10}}: \quad \beta^{\SU(4)} & = -3 \times 4 + 24 \times \frac{1}{2} + 2 \times 1 = 2 \,, \nonumber  \\
	\text{Model~\hyperref[spec:r19]{r19}}: \quad \beta^{\SU(4)} & = -3 \times 4 + 16 \times \frac{1}{2} + 2 \times 1 = -2 \,, \nonumber \\
	\text{Model~\hyperref[spec:r20]{r20}}: \quad \beta^{\SU(4)} & = -3 \times 4 + 16 \times \frac{1}{2} + 2 \times 1 = -2 \,, \nonumber \\
	\text{Model~\hyperref[spec:r22]{r22}}: \quad \beta^{\SU(4)} & = -3 \times 4 + 24 \times \frac{1}{2} + 2 \times 1 = 2 \,, \nonumber  \\
	\text{Model~\hyperref[spec:r25]{r25}}: \quad \beta^{\SU(4)} & = -3 \times 4 + 12 \times \frac{1}{2} + 2 \times 1 = -4 \,, \nonumber \\
	\text{Model~\hyperref[spec:r26]{r26}}: \quad \beta^{\SU(4)} & = -3 \times 4 + 16 \times \frac{1}{2} + 2 \times 1 = -2 \,, \nonumber \\
	\text{Model~\hyperref[spec:r29]{r29}}: \quad \beta^{\SU(4)} & = -3 \times 4 + 24 \times \frac{1}{2} + 2 \times 1 = 2 \,, \nonumber  \\
	\text{Model~\hyperref[spec:r30]{r30}}: \quad \beta^{\SU(4)} & = -3 \times 4 + 24 \times \frac{1}{2} + 2 \times 1 = 2 \,.            
\end{align}
From these results, we observe that models \hyperref[spec:r19]{r19}, \hyperref[spec:r20]{r20}, \hyperref[spec:r25]{r25}, and \hyperref[spec:r26]{r26} yield negative beta function coefficients, indicating that the corresponding $\SU(4)$ gauge groups are asymptotically free. In contrast, the remaining models have positive beta functions and thus do not exhibit asymptotic freedom.

\section{Conclusion}\label{sec:conclusion}

We have presented a class of consistent three-family $\mathcal{N}=1$ supersymmetric Pati-Salam models constructed from rigid intersecting D6-branes on the factorizable $\mathbb{T}^6/(\mathbb{Z}_2 \times \mathbb{Z}_2')$ orientifold with discrete torsion. These models are the first of their kind to be built entirely on a factorizable compactification lattice consisting of rectangular two-tori with frozen moduli. A key feature of the setup is the use of rigid three-cycles, which stabilize the open-string moduli associated with D-brane positions and Wilson lines. This eliminates unwanted adjoint matter and allows for negative beta functions in certain models, thereby ensuring asymptotic freedom. All models contain suitable GUT Higgs fields to ensure that the Pati–Salam gauge symmetry can be spontaneously broken down to that of the Standard Model.

The models satisfy the full set of consistency conditions, including $\mathcal{N}=1$ supersymmetry, K-theory constraints, RR tadpole cancellation, and the recently proposed swampland bounds on the maximal rank of the gauge group. We have provided detailed chiral spectra and discussed the decoupling of exotic states via strong dynamics in the hidden sector. A complete phenomenological analysis, including the computation of Yukawa couplings and the structure of soft supersymmetry breaking terms, is left for future work.

It would be interesting to explore variations of this model-building framework that exclude the $d$-stack of rigid D6-branes, while still enabling the construction of GUT Higgs fields responsible for breaking the Pati-Salam gauge symmetry. Such constructions may yield Standard Model-like spectra after appropriate symmetry breaking. Although our analysis has been confined to models with rectangular tori, the framework presented in Section~\ref{sec:model-building} remains applicable to tilted tori as well. However, to date, no three-family model with tilted tori and suitable GUT Higgs fields has been found within this framework. It would be valuable to investigate the origin of this obstruction more thoroughly, or ideally to identify a rigorous argument that rules out such constructions altogether.

A particularly compelling direction for future work involves a statistical exploration of the landscape of such vacua in string theory. In this context, it would be highly interesting to perform a systematic classification of all possible vacua using modern machine learning techniques, along the lines of~\cite{Ishiguro:2023hcv}.

\acknowledgments{We are grateful to the anonymous referee for the valuable suggestions that have greatly improved the presentation of the paper. MS is grateful to Gary Shiu for valuable discussion during the Lotus and the Swampland workshop. AM is supported by the Guangdong Basic and Applied Basic Research Foundation (Grant No. 2021B1515130007), Shenzhen Natural Science Fund (the Stable Support Plan Program 20220810130956001). TL is supported in part by the National Key Research and Development Program of China Grant No. 2020YFC2201504, by the Projects No. 11875062, No. 11947302, No. 12047503, and No. 12275333 supported by the National Natural Science Foundation of China, by the Key Research Program of the Chinese Academy of Sciences, Grant No. XDPB15, by the Scientific Instrument Developing Project of the Chinese Academy of Sciences, Grant No. YJKYYQ20190049, and by the International Partnership Program of Chinese Academy of Sciences for Grand Challenges, Grant No. 112311KYSB20210012. MS is supported in part by the National Natural Science Foundation of China (Grant No. 12475105).}

\appendix

\section{Three-family Pati-Salam models on rigid cycles}\label{appA} 

In this appendix, we present a list of the $10$ independent three-family $\mathcal{N}=1$ supersymmetric Pati-Salam models that arise from rigid intersecting D6-branes on a type IIA $\mathbb{T}^6/(\mathbb{Z}_2 \times \mathbb{Z}_2')$ orientifold. The models are named according to the rank of their gauge group and are sorted in ascending order. Additionally, we provide the relevant two-torus complex structure moduli and the gauge coupling relations in the captions.

\begin{table}[t]
	\centering
	$\begin{array}{|c|c|c|}
		\hline
		\text{Model~\hyperref[spec:r06]{r06}} & N & (n^1, m^1) \times (n^2, m^2) \times (n^3, m^3) \\
		\hline 
		a                                     & 4 & (0, 1)\times (0, -1)\times (1, 0)              \\
		b                                     & 2 & (1, 1)\times (-4, -1)\times (2, 1)             \\
		c                                     & 2 & (-4, -1)\times (0, -1)\times (-4, -1)          \\
		d                                     & 2 & (3, -1)\times (4, -1)\times (0, 1)             \\
		\hline
	\end{array}$
	\caption{Model~\hyperref[spec:r06]{r06} with the gauge group $\SU(4)_{C}\times\SU(2)_{L}\times\SU(2)_{R_1}\times\SU(2)_{R_2}$, the torus moduli $\chi_1=\sqrt{13}$, $\chi_2=\frac{12}{\sqrt{13}}$, $\chi_3= \frac{16}{\sqrt{13}}$ and the gauge coupling relation $g_a^2=\frac{308}{39}g_b^2=\frac{700}{39}g_R^2=\frac{51040}{27241}\frac{5 g_Y^2}{3}=\frac{16 \, \pi \,  e^{\phi _4}}{\sqrt{3}\sqrt[4]{13}}$.}
	\label{model:r06}
\end{table}

\begin{table}[t]
	\centering
	$\begin{array}{|c|c|c|}
		\hline
		\text{Model~\hyperref[spec:r08]{r08}} & N & (n^1, m^1) \times (n^2, m^2) \times (n^3, m^3) \\
		\hline 
		a                                     & 4 & (0, 1)\times (0, -1)\times (1, 0)              \\
		b                                     & 2 & (-1, -1)\times (4, 1)\times (2, 1)             \\
		c                                     & 2 & (-2, 1)\times (0, 1)\times (-4, 1)             \\
		d                                     & 2 & (3, -1)\times (-4, 1)\times (0, -1)            \\
		e_1                                   & 2 & (1, 0)\times (0, -1)\times (0, 1)              \\
		e_2                                   & 2 & (-1, 0)\times (0, 1)\times (0, 1)              \\
		\hline
	\end{array}$
	\caption{Model~\hyperref[spec:r08]{r08} with the gauge group $\SU(4)_{C}\times\SU(2)_{L}\times\SU(2)_{R_1}\times\SU(2)_{R_2}\times\SU(2)^2$, the torus moduli $\chi_1=\sqrt{5}$, $\chi_2=\frac{12}{\sqrt{5}}$, $\chi_3= \frac{8}{\sqrt{5}}$ and the gauge coupling relation $g_a^2=\frac{28}{5}g_b^2=\frac{44}{3}g_R^2=\frac{1680}{947}\frac{5 g_Y^2}{3}=\frac{8 \sqrt{\frac{2}{3}} \, \pi \,  e^{\phi _4}}{\sqrt[4]{5}}$.}
	\label{model:r08}
\end{table}

\begin{table}[t]
	\centering
	$\begin{array}{|c|c|c|}
		\hline
		\text{Model~\hyperref[spec:r10]{r10}} & N & (n^1, m^1) \times (n^2, m^2) \times (n^3, m^3) \\
		\hline 
		a                                     & 4 & (0, 1)\times (0, -1)\times (1, 0)              \\
		b                                     & 2 & (4, -1)\times (-1, 0)\times (-1, -1)           \\
		c                                     & 2 & (-4, 1)\times (3, -2)\times (1, -1)            \\
		d                                     & 2 & (0, 1)\times (-2, 1)\times (-2, 1)             \\
		e_1                                   & 2 & (0, -1)\times (1, 0)\times (0, 1)              \\
		e_2                                   & 2 & (0, 1)\times (-1, 0)\times (0, 1)              \\
		f_1                                   & 2 & (0, -1)\times (0, 1)\times (1, 0)              \\
		f_2                                   & 2 & (0, -1)\times (0, -1)\times (-1, 0)            \\
		\hline
	\end{array}$
	\caption{Model~\hyperref[spec:r10]{r10} with the gauge group $\SU(4)_{C}\times\SU(2)_{L}\times\SU(2)_{R_1}\times\SU(2)_{R_2}\times\SU(2)^4$, the torus moduli $\chi_1=8$, $\chi_2=2$, $\chi_3= 2$ and the gauge coupling relation $g_a^2=\frac{5}{4}g_b^2=\frac{41}{4}g_R^2=\frac{500}{323}\frac{5 g_Y^2}{3}=2 \sqrt{2} \, \pi \,  e^{\phi _4}$.}
	\label{model:r10}
\end{table}

\begin{table}[t]
	\centering
	$\begin{array}{|c|c|c|}
		\hline
		\text{Model~\hyperref[spec:r19]{r19}} & N & (n^1, m^1) \times (n^2, m^2) \times (n^3, m^3) \\
		\hline 
		a                                     & 4 & (0, 1)\times (0, -1)\times (1, 0)              \\
		b                                     & 2 & (-3, -1)\times (0, -1)\times (-2, -1)          \\
		c                                     & 2 & (2, 1)\times (-3, -1)\times (1, 1)             \\
		d                                     & 2 & (1, 1)\times (3, 1)\times (-2, -1)             \\
		e                                     & 2 & (0, 1)\times (0, -1)\times (1, 0)              \\
		f_1                                   & 2 & (1, 0)\times (0, -1)\times (0, 1)              \\
		f_2                                   & 2 & (-1, 0)\times (0, 1)\times (0, 1)              \\
		g_1                                   & 2 & (0, -1)\times (1, 0)\times (0, 1)              \\
		g_2                                   & 2 & (0, 1)\times (-1, 0)\times (0, 1)              \\
		h_1                                   & 2 & (1, 0)\times (1, 0)\times (1, 0)               \\
		h_2                                   & 2 & (-1, 0)\times (-1, 0)\times (1, 0)             \\
		h_3                                   & 2 & (1, 0)\times (-1, 0)\times (-1, 0)             \\
		h_4                                   & 2 & (-1, 0)\times (1, 0)\times (-1, 0)             \\
		\hline
	\end{array}$
	\caption{Model~\hyperref[spec:r19]{r19} with the gauge group $\SU(4)_{C}\times\SU(2)_{L}\times\SU(2)_{R_1}\times\SU(2)_{R_2}\times\SU(2)^5\times\USp(4)^4$, the torus moduli $\chi_1=\sqrt{6}$, $\chi_2=\frac{9 \sqrt{\frac{3}{2}}}{2}$, $\chi_3= \sqrt{6}$ and the gauge coupling relation $g_a^2=5g_b^2=\frac{70}{9}g_R^2=\frac{175}{124}\frac{5 g_Y^2}{3}=\frac{8\ 2^{3/4} \, \pi \,  e^{\phi _4}}{3 \sqrt[4]{3}}$.}
	\label{model:r19}
\end{table}

\begin{table}[t]
	\centering
	$\begin{array}{|c|c|c|}
		\hline
		\text{Model~\hyperref[spec:r20]{r20}} & N & (n^1, m^1) \times (n^2, m^2) \times (n^3, m^3) \\
		\hline 
		a                                     & 4 & (1, 0)\times (0, 1)\times (0, -1)              \\
		b                                     & 2 & (-1, -1)\times (3, 1)\times (0, 1)             \\
		c                                     & 2 & (-2, -1)\times (-1, -1)\times (-3, -1)         \\
		d                                     & 2 & (1, -1)\times (-2, 1)\times (3, -1)            \\
		e_1                                   & 2 & (1, 0)\times (0, -1)\times (0, 1)              \\
		e_2                                   & 2 & (-1, 0)\times (0, 1)\times (0, 1)              \\
		f_1                                   & 2 & (0, -1)\times (1, 0)\times (0, 1)              \\
		f_2                                   & 2 & (0, 1)\times (-1, 0)\times (0, 1)              \\
		g_1                                   & 2 & (0, -1)\times (0, 1)\times (1, 0)              \\
		g_2                                   & 2 & (0, -1)\times (0, -1)\times (-1, 0)            \\
		h_1                                   & 2 & (1, 0)\times (1, 0)\times (1, 0)               \\
		h_2                                   & 2 & (-1, 0)\times (-1, 0)\times (1, 0)             \\
		h_3                                   & 2 & (1, 0)\times (-1, 0)\times (-1, 0)             \\
		h_4                                   & 2 & (-1, 0)\times (1, 0)\times (-1, 0)             \\
		\hline
	\end{array}$
	\caption{Model~\hyperref[spec:r20]{r20} with the gauge group $\SU(4)_{C}\times\SU(2)_{L}\times\SU(2)_{R_1}\times\SU(2)_{R_2}\times\SU(2)^6\times\USp(4)^4$, the torus moduli $\chi_1=\sqrt{3}$, $\chi_2=\sqrt{3}$, $\chi_3= 9 \sqrt{3}$ and the gauge coupling relation $g_a^2=4g_b^2=\frac{56}{9}g_R^2=\frac{14}{11}\frac{5 g_Y^2}{3}=\frac{8 \, \pi \,  e^{\phi _4}}{3 \sqrt[4]{3}}$.}
	\label{model:r20}
\end{table}

\begin{table}[t]
	\centering
	$\begin{array}{|c|c|c|}
		\hline
		\text{Model~\hyperref[spec:r22]{r22}} & N & (n^1, m^1) \times (n^2, m^2) \times (n^3, m^3) \\
		\hline 
		a                                     & 4 & (0, 1)\times (0, -1)\times (1, 0)              \\
		b                                     & 2 & (-1, 0)\times (4, -1)\times (-2, -1)           \\
		c                                     & 2 & (4, -1)\times (0, 1)\times (2, -1)             \\
		d                                     & 2 & (-3, 2)\times (-4, 1)\times (-2, 1)            \\
		e_1                                   & 4 & (1, 0)\times (1, 0)\times (1, 0)               \\
		e_2                                   & 4 & (-1, 0)\times (-1, 0)\times (1, 0)             \\
		e_3                                   & 4 & (1, 0)\times (-1, 0)\times (-1, 0)             \\
		e_4                                   & 4 & (-1, 0)\times (1, 0)\times (-1, 0)             \\
		\hline
	\end{array}$
	\caption{Model~\hyperref[spec:r22]{r22} with the gauge group $\SU(4)_{C}\times\SU(2)_{L}\times\SU(2)_{R_1}\times\SU(2)_{R_2}\times\USp(8)^4$, the torus moduli $\chi_1=2$, $\chi_2=8$, $\chi_3= 4$ and the gauge coupling relation $g_a^2=\frac{5}{2}g_b^2=\frac{45}{2}g_R^2=\frac{250}{127}\frac{5 g_Y^2}{3}=4 \, \pi \,  e^{\phi _4}$.}
	\label{model:r22}
\end{table}

\begin{table}[t]
	\centering
	$\begin{array}{|c|c|c|}
		\hline
		\text{Model~\hyperref[spec:r25]{r25}} & N & (n^1, m^1) \times (n^2, m^2) \times (n^3, m^3) \\
		\hline 
		a                                     & 4 & (1, 0)\times (0, 1)\times (0, -1)              \\
		b                                     & 2 & (-2, -1)\times (-4, 1)\times (1, -1)           \\
		c                                     & 2 & (4, 1)\times (1, 1)\times (-3, -1)             \\
		d                                     & 2 & (-4, 1)\times (3, -1)\times (1, -1)            \\
		e                                     & 2 & (0, 1)\times (0, -1)\times (1, 0)              \\
		f_1                                   & 2 & (0, -1)\times (0, 1)\times (1, 0)              \\
		f_2                                   & 2 & (0, -1)\times (0, -1)\times (-1, 0)            \\
		g_1                                   & 4 & (1, 0)\times (1, 0)\times (1, 0)               \\
		g_2                                   & 4 & (-1, 0)\times (-1, 0)\times (1, 0)             \\
		g_3                                   & 4 & (1, 0)\times (-1, 0)\times (-1, 0)             \\
		g_4                                   & 4 & (-1, 0)\times (1, 0)\times (-1, 0)             \\
		\hline
	\end{array}$
	\caption{Model~\hyperref[spec:r25]{r25} with the gauge group $\SU(4)_{C}\times\SU(2)_{L}\times\SU(2)_{R_1}\times\SU(2)_{R_2}\times\SU(2)^3\times\USp(8)^4$, the torus moduli $\chi_1=2 \sqrt{611}$, $\chi_2=\sqrt{\frac{47}{13}}$, $\chi_3= \sqrt{\frac{47}{13}}$ and the gauge coupling relation $g_a^2=\frac{6120}{47}g_b^2=\frac{9840}{47}g_R^2=\frac{4100}{1687}\frac{5 g_Y^2}{3}=\frac{8 \sqrt{2} 13^{3/4} \, \pi \,  e^{\phi _4}}{\sqrt[4]{47}}$.}
	\label{model:r25}
\end{table}

\begin{table}[t]
	\centering
	$\begin{array}{|c|c|c|}
		\hline
		\text{Model~\hyperref[spec:r26]{r26}} & N & (n^1, m^1) \times (n^2, m^2) \times (n^3, m^3) \\
		\hline 
		a                                     & 4 & (0, 1)\times (1, 0)\times (0, -1)              \\
		b                                     & 2 & (-3, -1)\times (2, 1)\times (3, 1)             \\
		c                                     & 2 & (1, 0)\times (-2, -1)\times (-3, 1)            \\
		d                                     & 2 & (1, -1)\times (-3, -1)\times (-1, 0)           \\
		e                                     & 2 & (1, 0)\times (1, 0)\times (1, 0)               \\
		f_1                                   & 4 & (1, 0)\times (0, -1)\times (0, 1)              \\
		f_2                                   & 4 & (-1, 0)\times (0, 1)\times (0, 1)              \\
		g_1                                   & 4 & (0, -1)\times (1, 0)\times (0, 1)              \\
		g_2                                   & 4 & (0, 1)\times (-1, 0)\times (0, 1)              \\
		h_1                                   & 4 & (0, -1)\times (0, 1)\times (1, 0)              \\
		h_2                                   & 4 & (0, -1)\times (0, -1)\times (-1, 0)            \\
		\hline
	\end{array}$
	\caption{Model~\hyperref[spec:r26]{r26} with the gauge group $\SU(4)_{C}\times\SU(2)_{L}\times\SU(2)_{R_1}\times\SU(2)_{R_2}\times\USp(4)\times\SU(4)^6$, the torus moduli $\chi_1=\frac{2 \sqrt{10}}{3}$, $\chi_2=2 \sqrt{10}$, $\chi_3= 3 \sqrt{10}$ and the gauge coupling relation $g_a^2=\frac{121}{10}g_b^2=\frac{247}{60}g_R^2=\frac{2695}{3548}\frac{5 g_Y^2}{3}=\frac{4\ 2^{3/4} \, \pi \,  e^{\phi _4}}{\sqrt[4]{5}}$.}
	\label{model:r26}
\end{table}

\begin{table}[t]
	\centering
	$\begin{array}{|c|c|c|}
		\hline
		\text{Model~\hyperref[spec:r29]{r29}} & N & (n^1, m^1) \times (n^2, m^2) \times (n^3, m^3) \\
		\hline 
		a                                     & 4 & (0, 1)\times (1, 0)\times (0, -1)              \\
		b                                     & 2 & (-4, -1)\times (-2, 1)\times (1, 0)            \\
		c                                     & 2 & (4, 1)\times (2, 1)\times (-3, -1)             \\
		d                                     & 2 & (-1, 0)\times (-2, 1)\times (2, 1)             \\
		e                                     & 2 & (1, 0)\times (0, 1)\times (0, -1)              \\
		f_1                                   & 2 & (1, 0)\times (0, -1)\times (0, 1)              \\
		f_2                                   & 2 & (-1, 0)\times (0, 1)\times (0, 1)              \\
		g_1                                   & 4 & (0, -1)\times (1, 0)\times (0, 1)              \\
		g_2                                   & 4 & (0, 1)\times (-1, 0)\times (0, 1)              \\
		h_1                                   & 4 & (0, -1)\times (0, 1)\times (1, 0)              \\
		h_2                                   & 4 & (0, -1)\times (0, -1)\times (-1, 0)            \\
		i_1                                   & 2 & (1, 0)\times (1, 0)\times (1, 0)               \\
		i_2                                   & 2 & (-1, 0)\times (-1, 0)\times (1, 0)             \\
		i_3                                   & 2 & (1, 0)\times (-1, 0)\times (-1, 0)             \\
		i_4                                   & 2 & (-1, 0)\times (1, 0)\times (-1, 0)             \\
		\hline
	\end{array}$
	\caption{Model~\hyperref[spec:r29]{r29} with the gauge group $\SU(4)_{C}\times\SU(2)_{L}\times\SU(2)_{R_1}\times\SU(2)_{R_2}\times\SU(2)^3\times\SU(4)^4\times\USp(4)^4$, the torus moduli $\chi_1=8$, $\chi_2=4$, $\chi_3= 4$ and the gauge coupling relation $g_a^2=\frac{5}{4}g_b^2=\frac{55}{8}g_R^2=\frac{125}{182}\frac{5 g_Y^2}{3}=2 \sqrt{2} \, \pi \,  e^{\phi _4}$.}
	\label{model:r29}
\end{table}

\begin{table}[t]
	\centering
	$\begin{array}{|c|c|c|}
		\hline
		\text{Model~\hyperref[spec:r30]{r30}} & N & (n^1, m^1) \times (n^2, m^2) \times (n^3, m^3) \\
		\hline 
		a                                     & 4 & (0, 1)\times (0, -1)\times (1, 0)              \\
		b                                     & 2 & (-1, -1)\times (4, 1)\times (2, 1)             \\
		c                                     & 2 & (4, -1)\times (0, -1)\times (-3, 1)            \\
		d                                     & 2 & (-3, 1)\times (-4, 1)\times (-1, 1)            \\
		e_1                                   & 6 & (1, 0)\times (1, 0)\times (1, 0)               \\
		e_2                                   & 6 & (-1, 0)\times (-1, 0)\times (1, 0)             \\
		e_3                                   & 6 & (1, 0)\times (-1, 0)\times (-1, 0)             \\
		e_4                                   & 6 & (-1, 0)\times (1, 0)\times (-1, 0)             \\
		\hline
	\end{array}$
	\caption{Model~\hyperref[spec:r30]{r30} with the gauge group $\SU(4)_{C}\times\SU(2)_{L}\times\SU(2)_{R_1}\times\SU(2)_{R_2}\times\USp(12)^4$, the torus moduli $\chi_1=3 \sqrt{\frac{7}{2}}$, $\chi_2=10 \sqrt{\frac{2}{7}}$, $\chi_3= 4 \sqrt{\frac{2}{7}}$ and the gauge coupling relation $g_a^2=\frac{26}{7}g_b^2=\frac{118}{15}g_R^2=\frac{18525}{13192}\frac{5 g_Y^2}{3}=\frac{8\ 2^{3/4}}{\sqrt[4]{7} \sqrt{15}} \, \pi \, e^{\phi_4}$.}
	\label{model:r30}
\end{table}

\bibliographystyle{JHEP}

\providecommand{\href}[2]{#2}\begingroup\raggedright\endgroup

\end{document}